\documentclass[3p,10pt]{elsarticle}

\makeatletter
\def\ps@pprintTitle{%
 \let\@oddhead\@empty
 \let\@evenhead\@empty
 \def\@oddfoot{\centerline{\thepage}}%
 \let\@evenfoot\@oddfoot}
\makeatother

\usepackage{amssymb}

\usepackage[utf8]{inputenc}
\usepackage[T1]{fontenc}

\usepackage{amsfonts}
\usepackage{amsmath}
\usepackage{amssymb}
\usepackage{amsthm}
\usepackage{array}
\usepackage{hyperref}  %
\usepackage{cleveref}
\usepackage[inline]{enumitem}
\usepackage{float}
\usepackage{graphicx}
\usepackage{latexsym}
\usepackage{siunitx}
\usepackage{subcaption}
\usepackage{url}
\usepackage{xcolor}

\usepackage{tikz}
\usetikzlibrary{positioning,shapes}

\usepackage{physics}

\usepackage{dirtytalk}

\usepackage[normalem]{ulem}

\usepackage{esint}

\usepackage{bm}

\renewcommand{\Re}{\mathrm{Re}}  %
\newcommand*{\codename}[1]{\texttt{#1}}

\newcommand{\cubismAMR}{\codename{CubismAMR}}

\newcommand{\RR}{\mathbb{R}}
\renewcommand*{\vec}[1]{\mathbf{#1}}
\newcommand{\mydiamond}{\rotatebox[origin=c]{45}{\tiny $\blacksquare$}}

\DeclareMathOperator{\CELU}{CELU}
\DeclareMathOperator{\SoftPlus}{SoftPlus}
\DeclareMathOperator{\LogUniform}{LogUniform}
\newcommand{\valpha}{\bm{\alpha}}
\newcommand{\vbeta}{\bm{\beta}}
\newcommand{\vmu}{\bm{\mu}}
\newcommand{\vomega}{\bm{\omega}}
\newcommand{\vsigma}{\bm{\sigma}}
\newcommand{\vSigma}{\bm{\Sigma}}
\newcommand{\vtheta}{\bm{\theta}}
\newcommand{\vh}{\vec{h}}  %
\renewcommand{\vu}{\vec{u}}  %
\newcommand{\vx}{\vec{x}}
\newcommand{\vy}{\vec{y}}
\newcommand{\vz}{\vec{z}}
\newcommand{\vw}{\vec{w}}
\newcommand{\vF}{\vec{F}}
\newcommand{\vFcyl}{\vec{F}_\text{cyl}}

\newcommand{\Decoder}{\mathcal{D}}
\newcommand{\Encoder}{\mathcal{E}}
\newcommand{\vEF}{\vec{f}}
\newcommand{\vq}{\vec{q}}

\usepackage{graphicx}

\begin{document}

\begin{frontmatter}

\title{Adaptive learning of effective dynamics: Adaptive real-time, online modeling for complex systems}

\address[ETH]{Computational Science and Engineering Laboratory, ETH Z\"urich, CH-8092, Switzerland}
\address[Harvard]{School of Engineering and Applied Sciences, 29 Oxford Street, Harvard University, Cambridge, MA 02138, USA}
\address[Stanford]{Department of Computer Science, 353 Serra Hall, Stanford University, CA 94035, USA}

\author[ETH]{Ivica Ki\v{c}i\'{c}}
\ead{kicici@ethz.ch}
\author[ETH,Harvard]{Pantelis R. Vlachas}
\ead{pvlachas@ethz.ch}
\author[ETH,Harvard]{Georgios Arampatzis}
\ead{garampat@ethz.ch}
\author[ETH,Harvard]{Michail Chatzimanolakis}
\ead{michaich@ethz.ch}
\author[Stanford]{Leonidas~Guibas}
\ead{guibas@cs.stanford.edu}
\author[Harvard]{Petros Koumoutsakos\corref{cor1}}
\ead{petros@seas.harvard.edu}
\cortext[cor1]{Corresponding author}

\begin{abstract}
Predictive simulations are essential for applications ranging from weather forecasting to material design. The veracity of these simulations hinges on their capacity to capture the effective system dynamics. Massively parallel simulations predict the systems dynamics by resolving all spatiotemporal scales, often at a cost that prevents experimentation. On the other hand, reduced order models are fast but often limited by the linearization of the system dynamics and the adopted heuristic closures. We propose a novel systematic framework that bridges large scale simulations and  reduced order models  to extract and forecast adaptively the effective dynamics (AdaLED) of multiscale systems.  
AdaLED employs an autoencoder to identify reduced-order representations of the system dynamics and an ensemble of probabilistic recurrent neural networks (RNNs) as the latent time-stepper.
The framework alternates between the computational solver and the surrogate, accelerating learned dynamics while leaving yet-to-be-learned dynamics regimes to the original solver.
AdaLED continuously adapts the surrogate to the new dynamics through online training.
The  transitions between the surrogate and the computational solver are determined by monitoring the prediction accuracy and uncertainty of the surrogate.
The effectiveness of AdaLED is demonstrated on three different systems - a Van der Pol oscillator, a 2D reaction-diffusion equation, and a 2D Navier-Stokes flow past a cylinder for varying Reynolds numbers (400 up to 1200), showcasing its ability to learn effective dynamics online, detect unseen dynamics regimes, and provide net speed-ups.
To the best of our knowledge, AdaLED is the first framework that couples a surrogate model with a computational solver to achieve online adaptive learning of effective dynamics.
It constitutes a potent tool for applications requiring many computationally expensive simulations.

\end{abstract}

\begin{keyword}
adaptive reduced-order modeling
\sep computer simulations
\sep machine learning
\sep online real-time learning
\sep continuous learning
\sep Navier-Stokes equations

\end{keyword}

\end{frontmatter}

\section{Introduction}
\label{sec:intro}

Simulations of complex systems  have transformed our predictive capabilities in areas ranging from health and epidemiology~\cite{lateef2010simulation} to physics~\cite{springel2005simulations}, meteorology~\cite{kurth2018exascale}, and fluid mechanics.
Large-scale, simulations are prominent in fields where experiments may be unavailable, such as astrophysics and climate sciences, or where they require expensive infrastructure, equipment, and personnel.

The predictive fidelity of the simulations depends on their capacity to resolve all relevant spatiotemporal scales of the physical phenomenon under study.
However, high fidelity implies high computational cost, which hinders experimentation and optimization.
Many scientific and engineering tasks, such as parameter and design optimization~\cite{ghattas_willcox_2021}, multi-objective optimization~\cite{gong2015multi}, reinforcement learning (RL)~\cite{verma2018efficient,novati2021automating,du2019good}, and high-throughput computing~\cite{Taufer2021}, require a large number of system evaluations.
While using highly accurate simulations for these tasks is desirable, it can also be cost-prohibitive and potentially infeasible.
As a result, numerous research efforts have focused on developing accurate and efficient surrogate models that can replicate and accelerate simulations.

While computationally costly simulations are essential in resolving all scales of a complex system, key quantities of interest can be often described by a coarse-grained, averaged behavior. Selecting the proper degrees of freedom for such coarse grained representations is a long standing problem in science and engineering. Furthermore, appropriate combinations  of coarse-grained and fine-scale simulations, predictions offer the potential for  accelerated simulations  at a controlled accuracy.
Pioneering hybrid methods include the Equation-Free Framework (EFF)~\cite{kevrekidis2003equation,laing2010reduced,bar2019learning}, the Heterogeneous Multiscale Method (HMM)~\cite{weinan2003heterognous,weinan2007heterogeneous}, and the FLow AVeraged integratoR (FLAVOR)~\cite{tao2010nonintrusive}.
Hybrid methods distinguish between a detailed high-dimensional physical space (\emph{micro scale}) which is expensive to simulate, and a coarse-grained, reduced-order, or latent space (\emph{macro scale}).
More specifically, in EFF, a system is first advanced in the expensive micro-scale for a given time.
Then, a transition is made into the macro scale using a compression mechanism, such as Principal Component Analysis, Dynamic Mode Decomposition~\cite{kutz2016dynamic}, or diffusion maps~\cite{coifman2006diffusion}.
EFF then employs time-stepping schemes such as Euler or Runge-Kutta to advance the macro-scale dynamics.
After several time steps, the macro-scale dynamics are mapped back onto the fine scale for detailed simulation.
By alternating between the micro-scale and the macro-scale dynamics at timescales of interest, EFF can achieve significant computational savings.
However, the generalization of EFF to complex high-dimensional systems has been limited by the proper information transfer between micro and macro and the use of inefficient  macro-scale propagator.

In recent years there have been numerous efforts to develop reduced-order models and accelerate complex simulations using machine learning (ML)~\cite{vlachas2020backpropagation,vlachas2018data,wan2018data,brunton2019machine,vinuesa2022enhancing,kochkov2021machine}.
In a previous works, we  extended the EFF with ML algorithms that learn the time integrators and the transfer operators in a data-driven manner.
The resulting framework of Learning the Effective Dynamics (LED)~\cite{vlachas2022multiscale} of complex dynamical systems employs convolutional autoencoders (CAEs) for the identification of the micro-to-macro and macro-to-micro mappings and recurrent neural networks (RNNs) to propagate the macro dynamics.
The autoencoder and the RNN are trained offline using data from the micro propagator, i.e., the original simulator.
LED has been applied to a variety of dynamical systems~\cite{vlachas2022learning}, from fluid flows to molecular simulations~\cite{vlachas2021accelerated}.

We note that frameworks related to LED include the Latent Evolution of Partial Differential Equations (LE-PDE)~\cite{wu2022learning}.
Meanwhile, CAEs coupled with Long Short-Term Memory networks (LSTMs) have been applied in modeling complex flows~\cite{wiewel2019latent, gonzalez2018deep, fukami2021model,stachenfeld2021learned,geneva2020modeling,maulik2021reduced,hasegawa2020machine}. Other autoencoders (AEs),coupled with LSTM networks have been employed for surrogate modeling of high-dimensional dynamical systems~\cite{pant2021deep}, e.g., unsteady flows over a circular cylinder~\cite{eivazi2020deep,zhang2022unsteady}, or structural modeling of a two-story building~\cite{simpson2021machine}. Other notable works are based  on Proper Orthogonal Decomposition (POD)~\cite{wu2020data,fresca2022pod,peherstorfer2016data,benner2017model,galbally2010non}, local approximations with POD~\cite{vlachas2021local, vlachas2022parametric}, Dynamic Mode Decomposition (DMD)~\cite{kutz2016dynamic}, and Dynamics Identification (ID)~\cite{fries2022lasdi,he2022glasdi}.
Adaptive extensions that utilize low-rank updates are proposed in~\cite{peherstorfer2015dynamic,peherstorfer2015online}.

However, to the best of our knowledge, the above mentioned frameworks do not entail one or more of the following characteristics:
\begin{enumerate}
    \item They lack continuous training, which limits their ability to adapt to changing dynamics or to generalize to regions underrepresented in the initial training data.
    \item They do not quantify prediction uncertainty or monitor prediction error.
    \item They overlook that a surrogate should only be used when it is reliable.
    \item They do not exploit the capability of restarting a computer simulation from any time point.
    \item They do not control the balance between speed-up and accuracy.
    \item They do not account for variations in parameters of the fine-scale dynamics.
\end{enumerate}
Although multiple works present robust surrogate modeling frameworks, the need for real-time applications and tasks involving non-stationary dynamics or state-space exploration calls for continual learning frameworks to address the problem of distribution shift~\cite{wang2022continual}.

Adaptivity and continual learning in forming surrogates are key components of the present paper.
The proposed method of \emph{Adaptive Learning of Effective Dynamics} (AdaLED) accelerates computationally expensive simulations by learning an online surrogate model that replaces the original simulator only when its predictions are sufficiently reliable. More specifically,
AdaLED extends the LED framework with uncertainty quantification.
The surrogate monitors its prediction accuracy and provides a confidence level for its predictions.
In turn, the surrogate is utilized only in state-space regions (parts of the trajectories) that it has learned and is confident about its prediction.
Otherwise, the computational solver is employed to simulate dynamics unknown to the surrogate model.
Finally, we extend LED with continuous learning capabilities to address the distribution shift problem in environments/dynamical systems with time-varying dynamics.

We demonstrate that AdaLED can adaptively learn complex dynamics, produce reliable surrogate model predictions, and accelerate computationally expensive simulations while maintaining high accuracy.
AdaLED provides  control over the accuracy-speed trade-offs by adaptively specifying error thresholds for the simulation.

The paper is organized as follows:
in \cref{sec:method}, we present a detailed description of the AdaLED framework while in 
\cref{sec:vdp,sec:rd}, we demonstrate the efficiency and efficacy of AdaLED on the Van der Pol oscillator and a 2D reaction-diffusion system, respectively.
In \cref{sec:cyl}, we show how AdaLED can accelerate a 2D Navier-Stokes simulation of flow past a cylinder at varying Reynolds numbers.
\Cref{sec:discussion} concludes the paper.
Technical information on the neural networks and handling of very high-dimensional fluid flow states are provided in the appendix.

\section{Method}
\label{sec:method}
We consider the evolution of a dynamical system at a micro/fine scale, with a state denoted by $\vx_t \in \RR^{d_x}$ at time $t$. 
The state is advanced by $\delta t$ using a \emph{micro propagator} $\mathcal{F}$, so that 
\begin{equation}
  \vx_{t+\delta t} = \mathcal{F}(\vx_t, \vEF_t)\text{.}
\end{equation}
where and by $\vEF_t \in \RR^{d_f}$ the time-varying external forcing that affects the dynamics. Examples of such system dynamics are Direct Numerical Simulations for turbulent flows and molecular dynamics in materials processes. 
The simulation provides access to quantities of interest denoted by $\vq_t \in \RR^{d_q}$, $\vq_t = \mathcal{Q}(\vx_t)$.
We postulate that the effective system dynamics can be approximated by lower-dimensional latent states $\vz_t \in \mathcal{M}_z$, where $\mathcal{M}_z \in \RR^{d_z}$ (with $d_z<<d_x$)is a low-order manifold of the system state space.
Here we  identify the latent space by  employing  an encoder $\Encoder^{\vtheta_\Encoder} : \RR^{d_x} \to \RR^{d_z}$ with trainable parameters $\vtheta_\Encoder$.
The encoder maps micro states $\vx_t$ to latent (macro) states $\vz_t = \Encoder^{\vtheta_\Encoder}(\vx_t)$.
In the other direction, a decoder $\Decoder^{\vtheta_\Decoder} : \RR^{d_z} \to \RR^{d_x}$, with trainable parameters $\vtheta_\Decoder$, maps the latent state $\vz_t$ to the micro state $\tilde{\vx}_t = \Decoder^{\vtheta_\Decoder}(\vz_t)$.
The optimal parameters $\vtheta_\Encoder^*$ and $\vtheta_\Decoder^*$ minimize an application-specific reconstruction loss $\ell(\vx_t, \tilde{\vx}_t)$:
\begin{equation}
  (\vtheta_\Encoder^*, \vtheta_\Decoder^*)
    = \operatorname*{arg\,min}_{\vtheta_\Encoder, \vtheta_\Decoder}
      \ell \left( \vx_t, \tilde{\vx}_t \right)
    = \operatorname*{arg\,min}_{\vtheta_\Encoder, \vtheta_\Decoder}
      \ell \left( \vx_t, \Decoder^{\vtheta_\Decoder}( \Encoder^{\vtheta_\Encoder}( \vx_t ) ) \right).
\end{equation}

A non-linear macro propagator $(\mathcal{H}^{\vtheta_M}, \mathcal{Z}^{\vtheta_M}, \mathcal{Q}^{\vtheta_M}, \mathcal{S}^{\vtheta_M})$, with parameters $\vtheta_M$ and an internal hidden state $\vh_t$ capturing non-Markovian effects, is trained to predict the system dynamics in the macro scale:
\begin{equation}
  \vh_{t+\Delta t} = \mathcal{H}^{\vtheta_M}(\vz_t, \vq_t, \vEF_t, \vh_t), \quad
  \tilde{\vz}_{t+\Delta t} = \vz_t + \mathcal{Z}^{\vtheta_M}(\vh_{t+\Delta t}), \quad
  \tilde{\vq}_{t+\Delta t} = \vq_t + \mathcal{Q}^{\vtheta_M}(\vh_{t+\Delta t}). \quad
\end{equation}
where $\Delta t$ is the time step of the macro propagator, with $\Delta t$ being an integer multiple of $\delta t$.
The macro propagator is trained with backpropagation through time~\cite{werbos1988generalization} to minimize the combined mean square error (MSE) loss
$\norm{\tilde{\vz}_{t+\Delta t} - \vz_{t+\Delta t}} + \norm{\tilde{\vq}_{t+\Delta t} - \vq_{t+\Delta t}}$.
Optional weights can be added to each loss component to control their relative importance.

We note that the present framework also predicts physical quantities of interest while evolving the latent space dynamics.
Such quantities of interest $\tilde{\vq}_{t+\Delta t}$ could be computed by reverting to the fine-scale representation of the system dynamics, i.e., $\tilde{\vq}_{t+\Delta t} = \mathcal{Q}(\Decoder^{\vtheta_\Decoder}(\tilde{\vz}_{t+\Delta t}))$.
However, this approach has two drawbacks: (i) it requires evaluation of the relatively expensive decoder $\Decoder^{\vtheta_\Decoder}$ (ii) the function $\mathcal{Q}$ might not be explicitly available.
Consequently, the macro propagator is trained to predict $\tilde{\vq}_{t+\Delta t}$ directly.
In addition, the macro propagator outputs the uncertainty $\sigma_{t+\Delta t} \in \RR$, i.e.
\begin{equation}
  \sigma_{t+\Delta t} = \mathcal{S}^{\vtheta_M}(\vh_{t+\Delta t}).
\end{equation}
The uncertainty is used to robustly control the transitions between the micro and the macro propagator, as explained later in the text.

To achieve significant acceleration of the simulations, the macro propagator operates with a time step $\Delta t \gg \delta t$.
Here, the encoder and decoder are the two halves of a convolutional autoencoder, and the macro propagator is an ensemble of probabilistic recurrent neural networks (PRNNs)(see  \cref{sec:uncertainty}).
We note that AdaLED can incorporate various encoders and decoders and accommodate any macro propagator that can estimate the uncertainty of its own predictions.
We will refer to the combination of the encoder, decoder, and propagator as the Machine-Learned Model (MLM).

\subsection{AdaLED cycle (inference)}

\begin{figure*}
  \includegraphics[width=\textwidth]{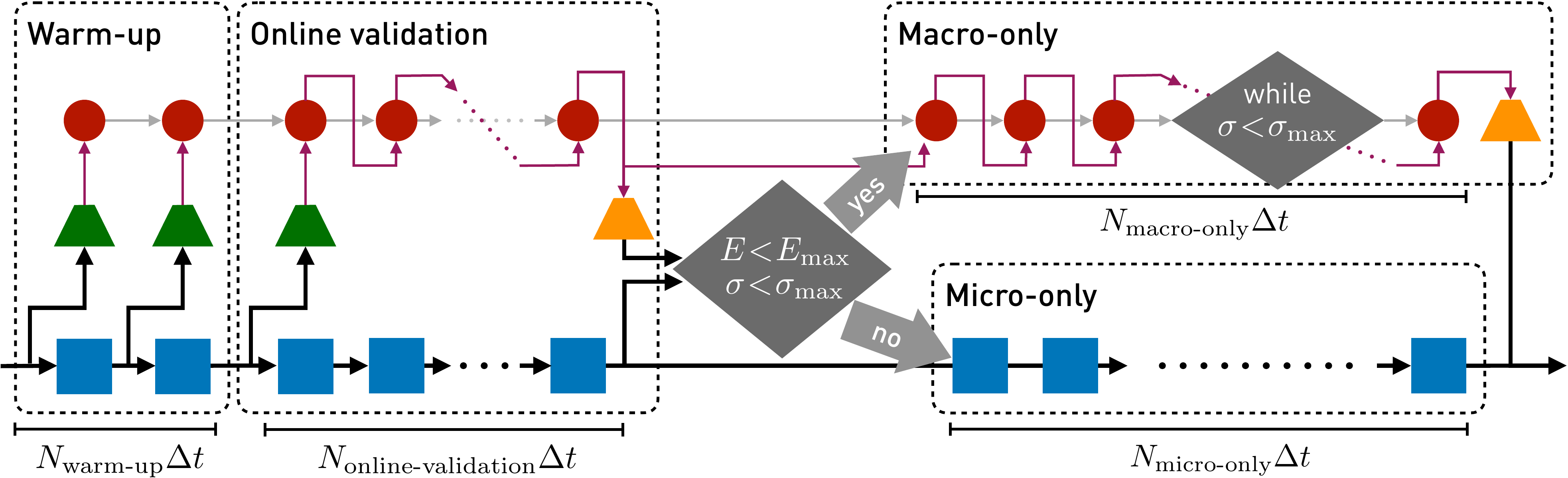}
  \caption{
    The stages of the inference, the AdaLED cycle.
    {\color[HTML]{0076BA} $\blacksquare$} denotes the micro propagator,
    {\large \color[HTML]{B51700} $\bullet$} the macro propagator,
    \includegraphics[height=0.8em]{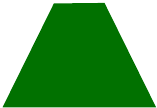} the encoder,
    \includegraphics[height=0.8em]{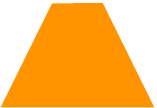} the decoder,
    \textbf{black} line the micro (high-dimensional) state $\vx_t$,
    {\color[HTML]{99195E} purple} the macro (latent) state $\vz_t$,
    and {\color{darkgray} gray} the hidden state $\vh_t$ of the macro propagator.
    Quantities of interest $\vq_t$ and external forcing $\vEF_t$ are hidden for brevity.
    Depending on the prediction error $E_t$ of the model and uncertainty $\sigma_t$ at the end of the online validation stage, either the macro-only or the micro-only stage is performed.
    The macro-only stage is performed as long as the uncertainty is below the threshold (possibly for 0 steps) or limited to a given number of steps.
  }
  \label{fig:adaled-cycle}
\end{figure*}
Inference in AdaLED proceeds in an iterative fashion.
In each iteration, AdaLED assesses the accuracy of the MLM and temporarily shifts the simulation from the micro to the macro scale if the accuracy is sufficiently high.
This alternation between the scales allows the micro propagator to correct errors introduced by the MLM and guide the simulation back to the manifold $\mathcal{M}_z$.
Additionally, the short cycles enable the MLM to replace the simulation in sections of trajectories that it has learned so far.
Other sections that are underrepresented in the training data or require more extended training are left to the micro propagator.
Finally, frequent evaluation of the micro propagator also enables continuous gathering of training data.

Each computational cycle in AdaLED consists of three stages: (i) the warm-up stage, (ii) the online validation stage, and (iii) either the micro-only or the macro-only stage~(\cref{fig:adaled-cycle}).
In the \emph{warm-up} stage, both micro and macro propagators are running.
In each time step, the micro state $\vx_t$ is passed through the encoder $\Encoder^{\vtheta_\Encoder}$ and fed into the macro propagator in order to warm up its hidden state $\vh_t$ (starting from $\vh_t = \vec{0}$).
In the \emph{online validation} stage, micro and macro propagators run independently, in order to estimate the macro propagator's prediction accuracy for the current section of the system trajectory.
At the end of the online validation stage, the final latent state $\tilde{\vz}_t$ is decoded back to the high-dimensional space, and an application-specific MLM prediction error $E_t = E(\vx_t, \Decoder^{\vtheta_\Decoder}(\tilde{\vz}_t))$ between the micro state (the ground truth) and the MLM's prediction is computed.
If either the MLM prediction error $E_t$ or the uncertainty $\sigma_t$ of the macro propagator are above the transition thresholds $E_\text{max}$ and $\sigma_\text{max}$, respectively, the macro prediction is discarded, and the simulation continues with the micro-only stage.
However, if both are below error thresholds, the prediction $\tilde{\vz}_t$ of the macro propagator is accepted, and the simulation continues with the macro-only stage.
Cycles are described as \emph{accepted} or \emph{rejected}, depending on whether the macro prediction was accepted or not.

In accepted cycles, the online validation stage is followed by the \emph{macro-only} stage, where the micro propagator is paused, and the only computation is done in the latent state using the inexpensive macro propagator.
This stage continues as long as the prediction uncertainty $\sigma_t$ is below the threshold $\sigma_\text{max}$.
Once the threshold is violated, the prediction for that step is dropped.
Then, the latent state from the previous time step is decoded to the high-dimensional state and passed to the micro propagator.
An important assumption is that the micro propagator can be reinitialized to an arbitrary state.
Additionally, this stage is optionally limited to $N^\text{max}_\text{macro-only}$ steps.
In \cref{fig:adaled-cycle}, values $N_X$ represent the number of time steps in the stage $X$.

\subsection{Dataset and training}

The trajectories $\vx_t$ produced by the micro propagator are sliced into trajectories of $L$ time steps and stored in a dynamic dataset of capacity $D \gg 1$.
The length $L$ is set equal to the number of recorded states in accepted cycles: $L = 1 + N_\text{warm-up} + N_\text{online-validation}$.
Once the dataset is filled, when adding a new trajectory, an existing trajectory selected uniformly at random is deleted.
Randomly removing trajectories ensures that old trajectories are preserved for a long time, alleviating the problem of catastrophic forgetting~\cite{kirkpatrick2017overcoming}, i.e., neural network predictions deteriorating in continuous learning for samples that they have seen in the past.

\begin{figure}
\centering
\includegraphics[width=0.6\columnwidth]{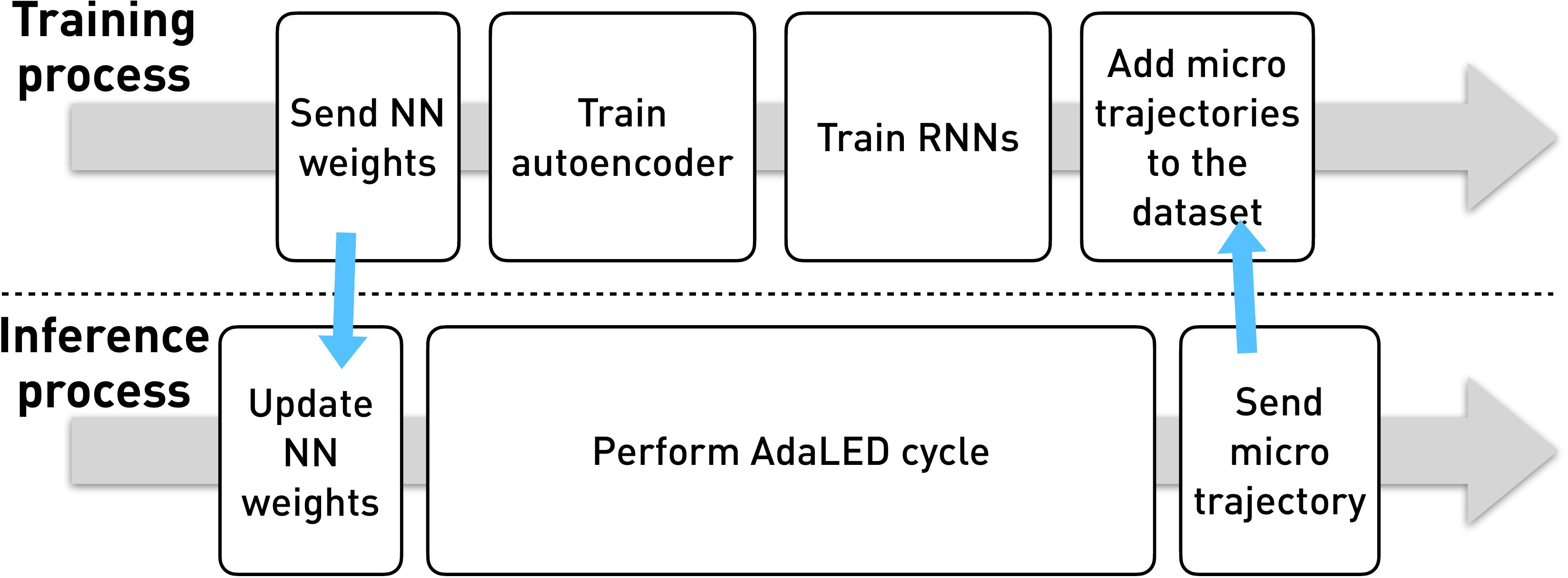}
\caption{
The inference and training loops.
For performance reasons, training and inference are optionally performed in separate processes.
Such division naturally extends to multiprocess training, multiprocess simulations, and to multiple simulations.
}
\label{fig:server-client}
\end{figure}

The autoencoder and the macro propagator are trained separately, one after the other, on a random subset of the dataset.
Training is performed continuously, either after each AdaLED cycle or asynchronously in parallel with AdaLED cycles~(\cref{fig:server-client}).
For inference, during one AdaLED cycle, the autoencoder and macro propagator parameters are fixed.

\subsection{Estimation of Prediction Uncertainty}
\label{sec:uncertainty}

In the  macro stage of AdaLED, the trajectory predicted by the macro propagator will eventually diverge from the ground truth , that the micro propagator would have produced, at a rate that depends on the complexity of the system dynamics \cite{vlachas2018data}.
In this study, rather than manually selecting the number of macro steps, we adopt a  robust mechanism for estimating the duration of reliable coarse-grained predictions.
To achieve this, we use probabilistic networks and network ensembles~\cite{lakshminarayanan2016simple}.

The input data are denoted by $\vx \in \RR^{N_x}$, and the output data (targets) of a network whose prediction uncertainty we want to estimate by $\vy \in \RR^{N_y}$.
The output of the system is predicted via two networks.
The first network is parameterized with~$\vtheta_{\vmu}$ and outputs the mean~$\vmu^{\vtheta_{\vmu}}(\vx)$.
This network is trained to minimize the MSE between the target and the mean output, i.e.,
\begin{equation}
\begin{aligned}
\ell_\text{MSE}(\vtheta_{\vmu}, \vx, \vy)
&= \frac{1}{N_y} \sum_{i=1}^{N_y} \left( \vmu^{\vtheta_{\vmu}}(\vx)_i - y_i \right)^2.
\end{aligned}
\end{equation}
A second network with parameters $\vtheta_{\vsigma}$ outputs the variance $\vSigma^{\vtheta_{\vsigma}}(\vx) = \operatorname{diag}(\vsigma^{\vtheta_{\vsigma}}(\vx)^2)$ of a Gaussian distribution with mean $\vmu^{\vtheta_{\vmu}}$, i.e., $p_{
\vtheta
}(\vy \vert \vx) = \mathcal{N}( \vy; \vmu^{\vtheta_{\vmu}}(\vx), \vSigma^{\vtheta_{\vsigma}}(\vx) )$~\cite{nix1994estimating, lakshminarayanan2016simple}, where $\vtheta=\{\vtheta_{\vmu},
\vtheta_{\vsigma}\}$.
A diagonal covariance matrix is considered here for simplicity.
The details of the neural architecture are shown in \cref{sec:lstm}.
This second network is trained to minimize the negative log-likelihood loss (NLL):
\begin{equation}
\begin{aligned}
\ell_\text{NLL}(\vtheta_{\vsigma}, \vx, \vy)
&= -\log p_{
\vtheta
}(\vy \vert \vx) \\
&= \frac{1}{2} \log \big(
\vsigma^{\vtheta_{\vsigma}}(\vx)
\big)^2
+ \frac{\left( \vy - \vmu^{\vtheta_{\vmu}}(\vx) \right)^2}{
2 \big(
\vsigma^{\vtheta_{\vsigma}}(\vx)
\big)^2
}
+ \text{const}.
\end{aligned}
\end{equation}
The networks are trained together, and can be viewed as a single network with parameters $\vtheta$, while  the weights $\vtheta_{\vmu}$ are considered fixed in the computation of the NLL loss.
The total sample loss can be written as:
\begin{equation}
\begin{aligned}
\ell(\vtheta, \vx, \vy)
&= \ell_\text{MSE}(\vtheta_{\vmu}, \vx, \vy) + \ell_\text{NLL}(\vtheta_{\vsigma}, \vx, \vy), \quad \vtheta=\{
\vtheta_{\vmu},
\vtheta_{\vsigma}
\}.
\end{aligned}
\end{equation}
This combination of MSE and NLL losses with decoupled gradients for $\vtheta_{\vmu}$ achieved higher accuracy than solely the NLL loss.

Moreover, we consider an ensemble of $K$ such probabilistic networks, each randomly initialized with its own parameters $\vtheta^k$, $k \in \{1, \dots, K\}$ and trained separately on the same data to minimize the loss $\ell(\vtheta^k, \vx, \vy)$.
For a given input $\vx$, the outputs $ \vmu^{(k)} = \vmu^{\vtheta_{\vmu}^k}(\vx)$ and $ \vsigma^{(k)}= \vsigma^{\vtheta_{\vsigma}^k}(\vx)$ are combined into the final prediction $\vmu(\vx)$ and uncertainty $\vsigma(\vx)$ of the ensemble as follows~\cite{lakshminarayanan2016simple}:
\begin{equation}
\label{eq:ensemble-mu-sigma}
\begin{aligned}
\vmu(\vx) &= \frac{1}{K} \sum_k \vmu^{(k)}(\vx)
\text{,} \\
\vsigma^2(\vx) &=
\underbrace{
\frac{1}{K} \sum_k
\big(\vsigma^{(k)}\big)^2(\vx)
}_{
\vsigma_{\text{ind}}^2(\vx)
}
+ \underbrace{
\frac{1}{K} \sum_k
\big( \vmu^{(k)} \big)^2 (\vx) -
\vmu^2(\vx)
}_{
\vsigma_{\text{std}}^2(\vx)
}.
\end{aligned}
\end{equation}
The term $\vsigma_\text{ind}$ in \cref{eq:ensemble-mu-sigma} refers to the prediction uncertainties as estimated by each network individually.
On the other hand, the term $\vsigma_\text{std}$ measures the disagreement of the ensemble in the prediction of $\vy$.
Thus, this term provides an estimate of the training inaccuracy for the given input $\vx$.
For unseen states $\vx$ or states underrepresented in the training data, we expected $\vsigma_\text{ind}$ and $\vsigma_\text{std}$ to be larger than for frequently seen states.
From the vector uncertainty $\vsigma$, the scalar uncertainty $\sigma$ is defined as either $\sigma = \norm{\vsigma}_2$ or $\sigma = \norm{\vsigma}_2 / N_y$.

\subsection{AdaLED hyper-parameters}

The error thresholds $E_\text{max}$ and $\sigma_\text{max}$ are application-specific and determine the trade-off between speed-up and accuracy.
The latent state dimension $d_z$ of the autoencoder should be chosen based on the system dynamics and its effective degrees of freedom~\cite{vlachas2022multiscale}.
The remaining hyperparameters, such as network size and the number of layers, can be determined through small-scale experiments and hyperparameter tuning.

\section{Case study: Van der Pol oscillator}
\label{sec:vdp}

\newcommand{\muA}{\mu_\text{ALT}}
\newcommand{\muB}{\mu_\text{RAND}}
\newcommand{\muC}{\mu_\text{BROWN}}

We first demonstrate the capabilities of AdaLED on the Van der Pol oscillator (VdP), a system used as a benchmark for a variety of multiscale frameworks~\cite{tao2010nonintrusive,kevrekidis2003equation,weinan2003heterognous}.
In contrast to these frameworks, we do not distinguish a priori between fast and slow dynamics.
Instead, we arbitrarily change the oscillator limit cycle and oscillation time scale, which is controlled by the damping parameter $\mu$, to demonstrate that AdaLED can adapt to these changes.
In this case study, no autoencoder is used, i.e., the encoder and the decoder are identity operators.

The Van der Pol oscillator~\cite{van1926lxxxviii, kaplan1997understanding} is a non-linear damped oscillator governed by the following equations:
\begin{equation}
\label{eq:vdp}
\begin{aligned}
\dv{x}{t} &= \mu \left( x - \frac{1}{3} x^3 - y \right)\text{,} \\
\dv{y}{t} &= \frac{1}{\mu} x\text{,}
\end{aligned}
\end{equation}
where $\mu = \mu(t) > 0$ is a time-varying system parameter that controls the system's non-linearity and damping.

The micro propagator is an ODE integrator based on the Euler method that integrates \cref{eq:vdp} with a time step of $\delta t = 0.001$ starting from a random initial condition $(x_0, y_0) \sim \mathcal{U}([-5, 5]^2)$.
The macro propagator is an ensemble of five LSTMs.
Their architecture is explained in~\cref{sec:lstm}.
The ensemble is trained to predict the dynamics with a macro time step of $\Delta t = 0.1$.
Each LSTM takes the tuple $(x(t), y(t), \mu(t))$ as input and outputs the means and variances $(\mu_{\Delta x}(t), \mu_{\Delta y}(t), \sigma^2_{\Delta x}(t), \sigma^2_{\Delta y}(t))$ for residuals $\Delta x(t) = x(t + \Delta t) - x(t)$ and $\Delta y(t) = y(t + \Delta t) - y(t)$.

For a constant $\mu$, the system enters a limit cycle whose shape depends on $\mu$~(\cref{fig:vdp-limit-cycles}).
For the values of $\mu$ and $\Delta t$ considered here, a single limit cycle is completed in about $60$ to $90$ macro time steps $\Delta t$.

AdaLED cycles are configured to have $7$ warm-up steps and $25$ online validation steps.
The maximum number of macro and micro steps is selected for each AdaLED cycle uniformly at random between $80$ and $120$ steps in order to avoid synchronizing the AdaLED cycle with the system limit cycle.
We observed that such synchronization causes the training dataset to be filled with data from the same limit cycle region while the rest of the cycle is underrepresented.
For a given cycle $c$, we define the \emph{macro utilization} $\eta_c = N^c_\text{macro-only}/N^c$ as the fraction of steps performed in the macro-only stage, where $N^c_\text{macro-only}$ is the number of macro-only steps and $N^c$ the total number of steps of the cycle $c$.
In rejected cycles, $\eta_c = 0$.
The total macro utilization $\eta$ is defined as $\eta = (\sum_c N^c_\text{macro-only})/(\sum_c N^c)$.
Given the selected stage durations, the maximum attainable total macro utilization is $\eta \approx 75.6\%$.

The maximum capacity of the dataset is set to $1280$ trajectories.
The training is performed at a fixed rate of 2 trajectories for each simulation macro step.

The prediction error is defined as $E = \sqrt{(x_\text{micro} - x_\text{macro})^2 + (y_\text{micro} - y_\text{macro})^2}$ and the prediction uncertainty as $\sigma = \sqrt{\sigma^2_{\Delta x} + \sigma^2_{\Delta y}}$.
The AdaLED transition thresholds are set to $E_\text{max} = 0.10$ and $\sigma_\text{max} = 0.10$.
We note that the error threshold $E_\text{max}$ refers to the accumulated error after 25 online validation steps and not a single-step prediction error.

\subsection{Results}

\begin{figure*}
  \begin{subfigure}{\textwidth}
    \centering
    \includegraphics[width=0.8\textwidth]{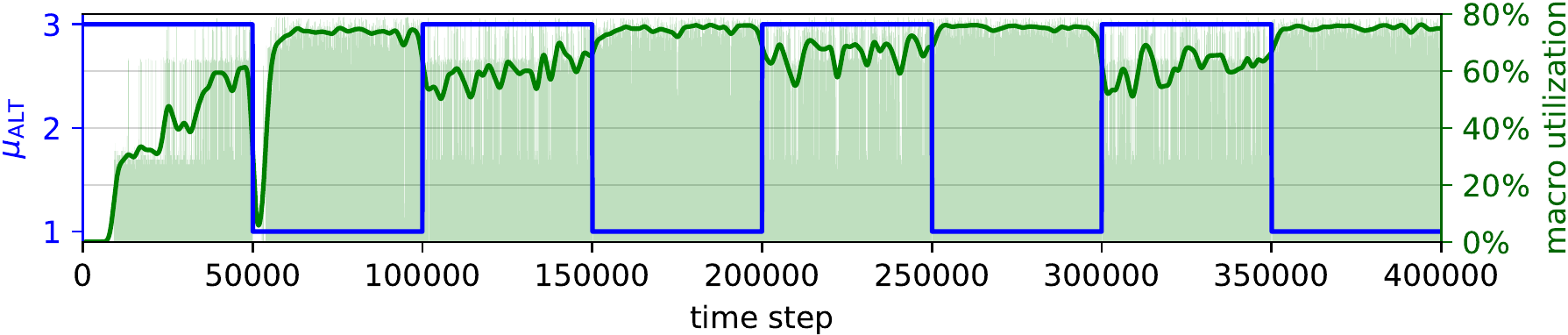}
  \end{subfigure}
  \medskip

  \begin{subfigure}{\textwidth}
    \centering
    \includegraphics[width=0.8\textwidth]{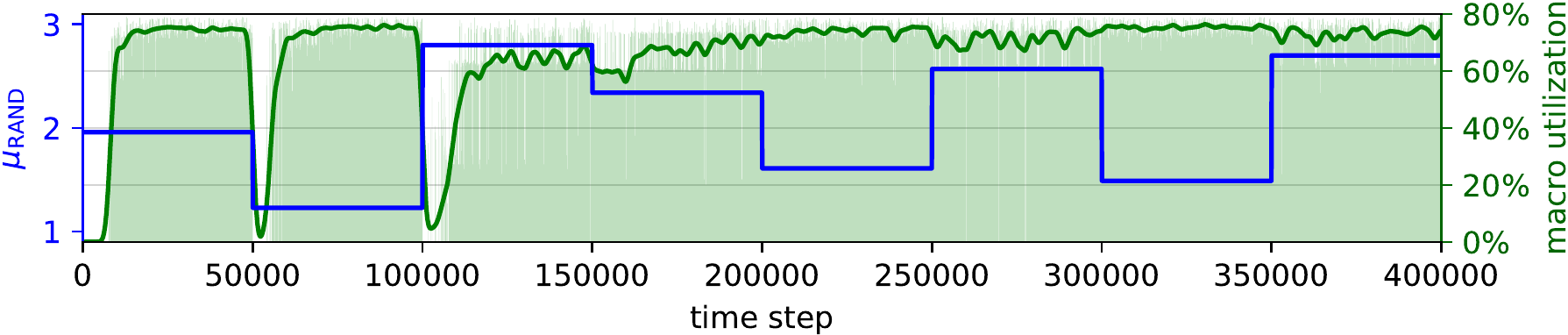}
  \end{subfigure}
  \medskip

  \begin{subfigure}{\textwidth}
    \centering
    \includegraphics[width=0.8\textwidth]{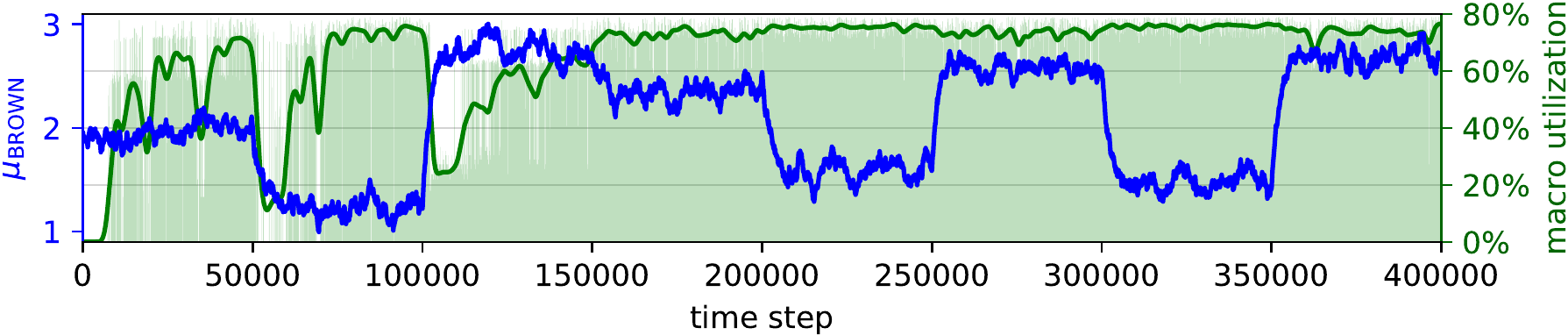}
  \end{subfigure}

  \caption{
    Macro utilization (fraction of steps performed in macro-only stage) for the Van der Pol oscillator case study for three different variants of $\mu(t)$.
    The light green histograms show the macro utilization of each individual AdaLED cycle, and the dark green line the macro utilization smoothed using a Gaussian blur (for visualization purposes only).
  }
  \label{fig:vdp-3-cases}
\end{figure*}

\begin{figure*}

  \begin{tikzpicture}
    \node[anchor=south west,inner sep=0] (image) at (0,0) {\includegraphics[width=\textwidth]{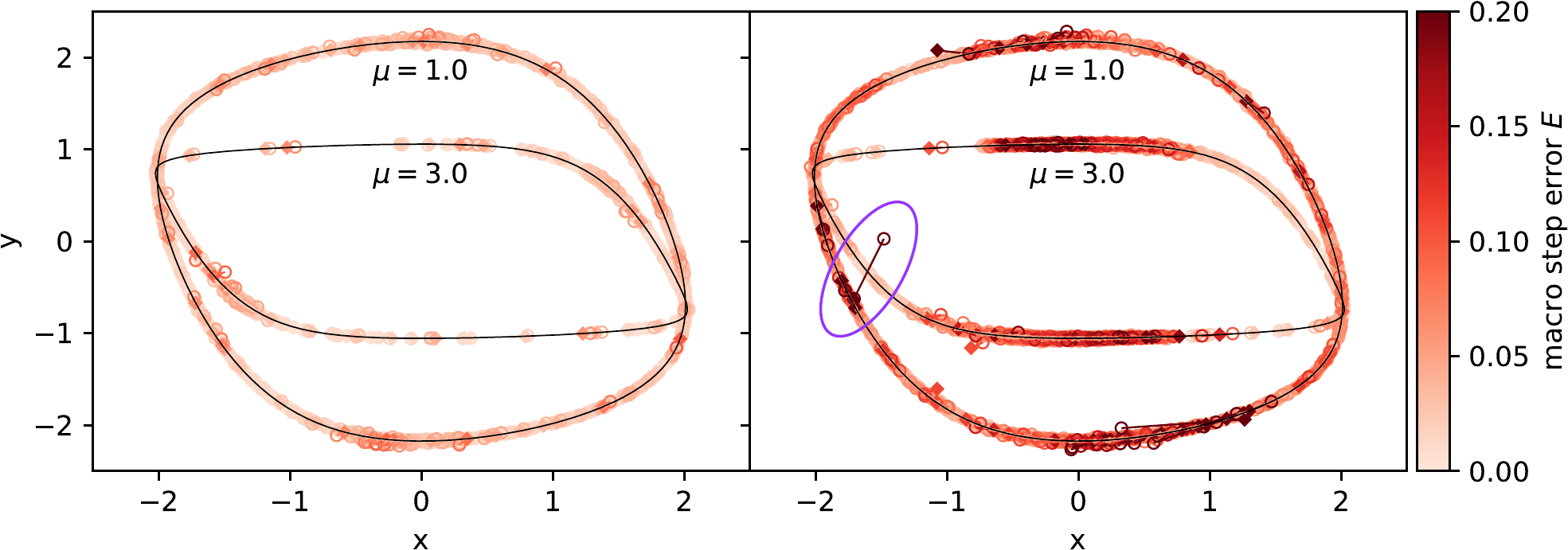}};
  \end{tikzpicture}

  \caption{
    Visualization of the prediction error for each accepted cycle in the case $\muA(t)$ of the Van der Pol oscillator case study.
    Left: error at the end of the online validation stages, right: error at the end of macro-only stages.
    Diamonds denote micro states and empty circles the macro states, colored with respect to the error $E$ (Euclidean distance).
    The AdaLED cycles with endpoints outside of limit cycles correspond to those during which the value of $\muA(t)$ changed.
    The first such change at the time step \num{50000} (encircled pair) is visualized in detail in \cref{fig:vdp-mu-change}.
  }
  \label{fig:vdp-mu1-validation}
\end{figure*}

\begin{figure}
  \centering
  \includegraphics[width=0.7\columnwidth]{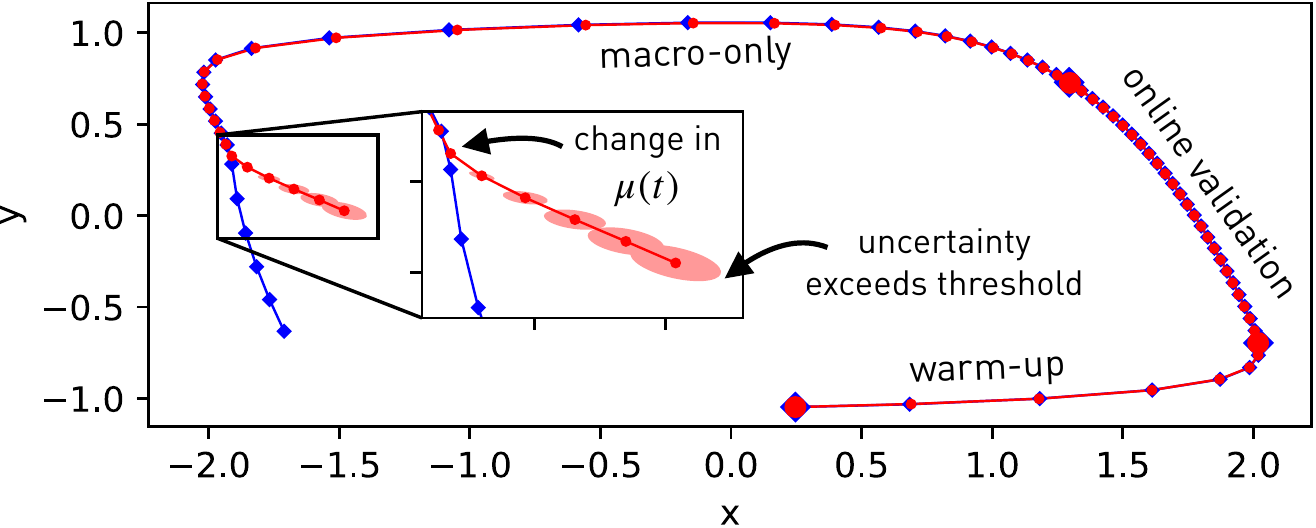}
  \caption{
    AdaLED cycle around the time step \num{50000} at which $\muA(t)$ switches from value $3.0$ to $1.0$.
    Blue diamonds {\color{blue}\mydiamond{}} denote the micro states, red circles {\color{red}$\bullet$} the macro states, and red ellipses the ensemble covariance.
    The plot shows that, due to the different responses of individual LSTMs in the ensemble to the value of $\mu(t)$, AdaLED quickly detected the change in dynamics and stopped the execution of the macro propagator.
  }
  \label{fig:vdp-mu-change}
\end{figure}

We analyze the performance of AdaLED on three different cases of $\mu(t)$:
\begin{itemize}
    \item $\muA(t)$, a piecewise constant function alternating between values $\mu = 1.0$ and $\mu = 3.0$ every \num{50000} time steps,
    \item $\muB(t)$, a piecewise constant function with multiple randomly selected values $\mu \in [1.0, 3.0]$, and
    \item $\muC(t)$, a piecewise constant function $\muB(t)$ augmented with Brownian-like noise (details explained in \cref{sec:appendix-vdp}).
\end{itemize}
The system is integrated for \num{400000} time steps.
The hyper-parameters of the LSTMs are tuned in a preliminary study reported in~\cref{sec:vdp-hyperparam-study}.

The macro utilization for all three cases, together with the functions $\mu(t)$, are shown in~\cref{fig:vdp-3-cases}.
The case $\muA(t)$ is depicted in the top figure.
The macro utilization at the start of the run is equal to zero, which is expected since the macro propagator is untrained and produces inaccurate predictions.
After about \num{8000} time steps, the prediction error reaches the desired threshold and AdaLED starts accepting the prediction of the macro propagator.
As a result, the macro utilization increases gradually to $60\%$.
At the time step \num{50000}, the value of $\mu(t)$ suddenly changes, putting the system into an unseen regime.
AdaLED correctly detects the change and starts rejecting the predictions of the macro propagator, which suddenly became unreliable.
However, soon after, the macro propagator learns the new regime and its predictions are accepted again, leading to an increase in macro utilization, which approaches the maximum of $75.6\%$.
After \num{50000} more time steps, the system switches back to $\mu=3$.
The macro propagator has already observed and learned this regime, as demonstrated by the fact that AdaLED resumes uninterruptedly with a very high acceptance rate.

The results for the case $\muB(t)$, shown in the middle plot of \cref{fig:vdp-3-cases}, display similar behavior as the previous case.
At the beginning of the simulation, all cycles are rejected until the network learns the corresponding regime.
This trend repeats after the first two changes in $\mu$.
From the third change onwards, we observe that AdaLED can interpolate between previously seen values of $\mu$ and can continue to produce accurate predictions despite changes on $\muB(t)$.
This demonstrates the ability of AdaLED to learn and interpolate on unseen dynamical regimes adaptively.

Finally, the more complex case of a noisy $\muC(t)$ is depicted in the bottom plot in \cref{fig:vdp-3-cases}.
AdaLED again gradually learns to replace the micro propagator and adaptively learns the different dynamical regimes.
However, here it takes longer for the macro propagator to reach the highest acceptance rate compared to the cases $\muA$ and $\muB$.
We argue that this is due to the increased difficulty of the learning task, as data are generated from various limit cycles with varying time scales.

The online validation errors and final testing errors (prediction error at the end of the macro-only stage) for each accepted cycle in case $\muA(t)$ are depicted in \cref{fig:vdp-mu1-validation}.
The testing error is calculated by running the micro propagator even during the macro-only stage, solely for evaluation purposes.
In production runs, the micro propagator is inactive during the macro-only stages.

We observe that errors are generally small and almost every cycle ends very close to the limit cycles.
Exceptions to this trend occur during sudden changes of $\mu(t)$.
It is important to note that the online validation errors (left plot) can be directly controlled by adjusting the value of $E_\text{max}$.
On the other hand, the testing error (right plot) can be controlled only indirectly through $E_\text{max}$ and $\sigma_\text{max}$, and thus serves as a measure of the robustness and quality of the surrogate model.
This highlights the advantage of AdaLED compared to other multiscale frameworks, as it allows for control over the error thresholds a priori.

The behavior of the system during the first change of $\muA$, happening at time step \num{50000}, is visualized in \cref{fig:vdp-mu-change}.
Prior to that change, the LSTMs were trained only on trajectories with $\mu = 3$.
As a result, the predictions of individual networks in the ensemble are alike and accurate for this particular regime.
However, once the value of $\muA(t)$ changes, different LSTMs in the ensemble respond differently to $\mu = 1$, since the predictions for the unseen regimes are arbitrary and depend on weight initialization.
This causes the predictions to diverge and the uncertainty to increase (as defined in \cref{eq:ensemble-mu-sigma}).
When the uncertainty crosses the threshold $\sigma_\text{max}$ after seven steps, the cycle terminates.
The 7th step is rejected and therefore excluded from the plot.

Additional results on the dependence of the testing error on the uncertainty threshold $\sigma_\text{max}$ and the ensemble size $K$ are presented in \cref{sec:vdp-E-vs-sigma-max-K}.

\section{Case study: Reaction-diffusion equation}
\label{sec:rd}

Here, we test AdaLED on the lambda-omega reaction-diffusion system~\cite{champion2019data,floryan2022data} governed by:
\begin{equation}
  \label{eq:rd}
  \begin{aligned}
    \pdv{u}{t} &= [1 - (u^2 + v^2)] u + \beta (u^2 + v^2) v + d_1 \nabla^2 u, \\
    \pdv{v}{t} &= -\beta (u^2 + v^2) u + [1 - (u^2 + v^2)] v + d_2 \nabla^2 v
  \end{aligned}
\end{equation}
for $-10 \leq x, y \leq 10$, where $\beta = 1.0$ is the reaction parameter and $d_1 = d_2 = d = d(t)$ the time-varying diffusion parameters.
The equation is integrated on a $96 \times 96$ uniform grid using the Runge–Kutta–Fehlberg method of fourth order with a time step of $\Delta t = 0.05$.
Thus, the state of the system is fully described by a tensor $\vec{w} = (\vec{u}, \vec{v}) \in \mathbb{R}^{2 \times 96 \times 96}$.
The system exhibits a spiral wave whose shape depends on the parameter $d$.

We evaluate the performance of AdaLED with the diffusion parameter $d$ alternating between $0.1$ and $0.2$ every \num{20000} time steps.
AdaLED cycles are configured to have 5 warm-up steps, 18 online validation steps, 10 to 15 micro-only steps, and 400 to 500 macro-only steps.
The AdaLED transition thresholds are set to $E_\text{max} = 0.002$ and $\sigma_\text{max} = 0.002$.
We use the mean square error (MSE) between the macro and the micro state $\vec{w}$ as the error metric, i.e., $E = E(\vec{w}, \tilde{\vec{w}}) = \norm{\vec{w} - \tilde{\vec{w}}}^2_2 / (2 \cdot 96 \cdot 96)$.
The simulation is run for \num{200000} time steps.
Other details of the system, the neural networks, and the training are listed in \cref{sec:appendix-rd}.

\begin{figure*}
  \centering
  \includegraphics[width=\textwidth]{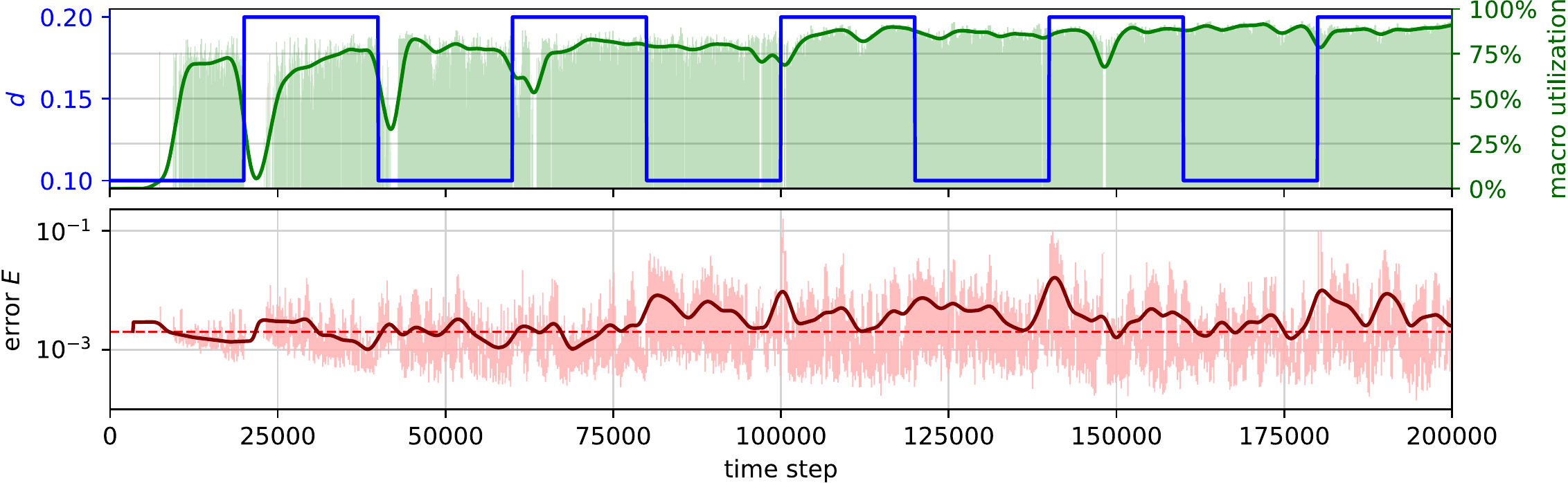}
  \caption{
    Performance on AdaLED on the reaction-diffusion case study with a time-varying diffusion parameter $d$ (blue line).
    Top: macro utilization (fraction of steps performed in macro-only stage).
    The light green histograms show the macro utilization of each individual AdaLED cycle, and the dark green line shows the macro utilization smoothed using a Gaussian blur (for visualization purposes only).
    Bottom: test error of the reconstructed state $\vec{w}$.
    The per-step errors (faded red) alternate between low values at the beginning of the macro-only stage and higher errors at the end of the macro-only stage.
    The dark red denotes the smoothed test error, and the dashed red the online validation threshold $E_\text{max}$.
  }
  \label{fig:rd-utilization-F-E}
\end{figure*}

The macro utilization $\eta$ and test errors $E$ are shown in \cref{fig:rd-utilization-F-E}.
Similar to the Van der Pol oscillator case study, we observe that the macro utilization is initially zero, as the networks are not yet trained.
However, after approximately \num{10000} time steps, the autoencoder learns to reconstruct the state, and the LSTMs learn to predict the dynamics.
Consequently, AdaLED begins to accept the macro prediction.
When the diffusion parameter $d$ changes from 0.1 to 0.2, AdaLED recognizes the unreliability of the macro predictions and switches back to the simulator.
Once the new regime is learned, AdaLED resumes using the macro propagator.
The bottom plot of \cref{fig:rd-utilization-F-E} shows the test errors $E$.
Overall, the macro utilization reaches $75\% \pm 1\%$, with an MSE of $0.0055 \pm 0.0013$ (relative MSE of $0.012 \pm 0.003$).
The reported confidence levels are based on the variance calculated from ten repeated simulation runs.

\begin{figure*}
  \centering
  \includegraphics[width=\textwidth]{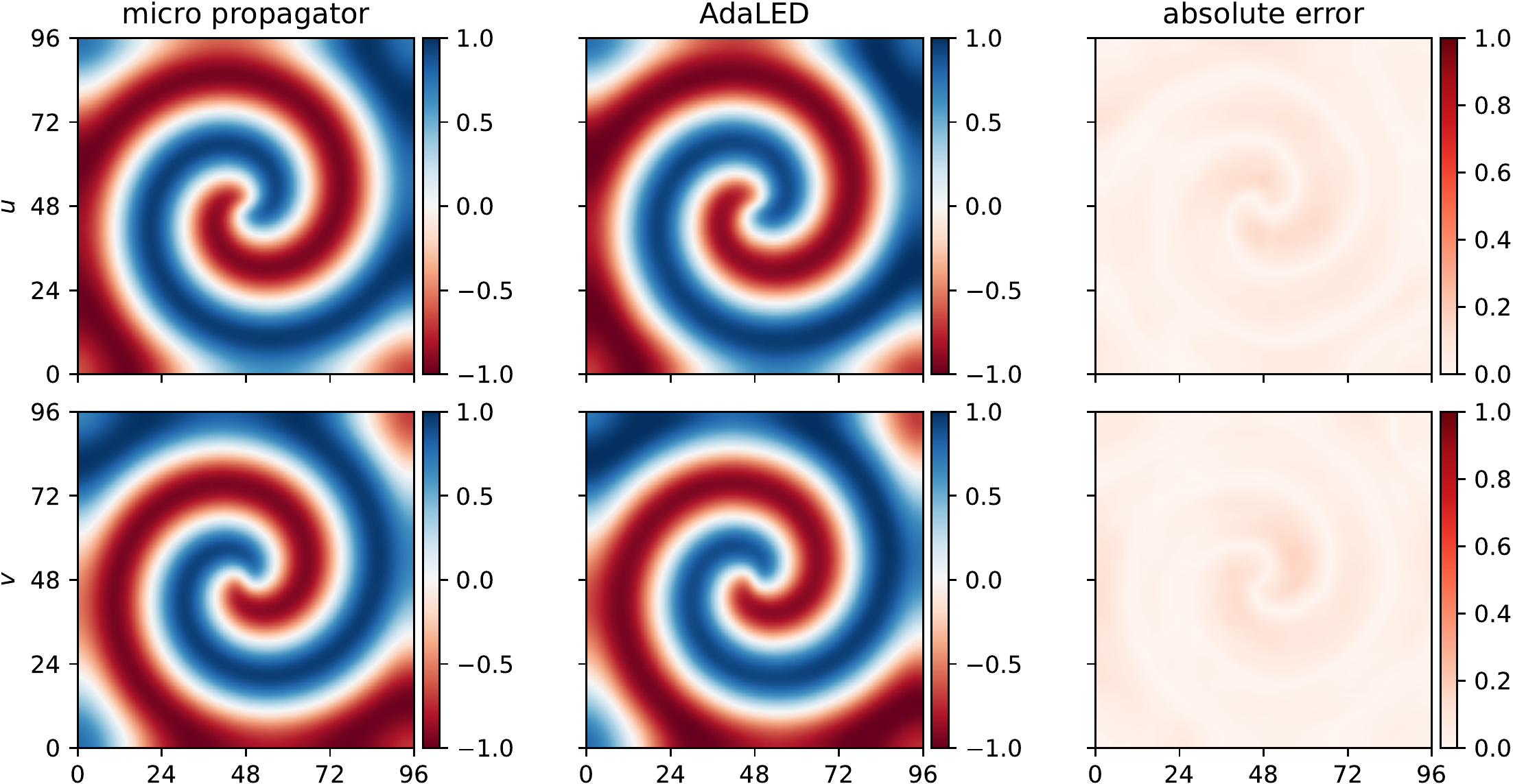}
  \caption{
    Snapshot of the simulation at time step $200000$, 305 steps into the macro-only stage.
    Left: the micro propagator (ground truth), middle: the AdaLED Machine-Learned Model, right: absolute error.
    The mean square error amounts to $E = 0.0021$ (relative error of $0.0045$).
  }
  \label{fig:rd-snapshot}
\end{figure*}

\Cref{fig:rd-snapshot} shows a snapshot of the simulation at the time step \num{200000}, 305 time steps into the macro-only stage.
The predicted and expected states are in agreement, demonstrating the accuracy of the macro propagator even after performing the simulation for a significant number of time steps in the latent space.
The micro states during macro-only stages are retrieved for testing purposes by continuing the micro simulation even during the macro-only stage.

\section{Case study: 2D flow past a cylinder}
\label{sec:cyl}

Finally, we employ AdaLED to accelerate a 2D Direct Numerical Simulation (DNS)  of the flow past a circular cylinder at  varying Reynolds numbers.
AdaLED is trained to forecast the velocity field, representing the state of the flow.
Forecasting the complete simulation state enables the alternation between macro (latent) and micro scale (the DNS ).
In addition to predicting the state, AdaLED is also tasked with predicting the force exerted by the fluid on the cylinder.
The force serves as the quantity of interest $\vq(t)$ that we want to have access to at all time steps of the simulation.

The system is governed by the incompressible Navier-Stokes equations and the no-slip boundary conidition is enforced via the Brinkman penalization~\cite{angot1999penalization}:
\begin{align}
  \label{eq:ns-incompressible}
  \nabla \cdot \vec{u} &= 0, \\
  \label{eq:ns-brinkman}
  \pdv{\vec{u}}{t} + (\vec{u} \cdot \nabla) \vec{u}
    &= -\frac{1}{\rho} \nabla p + \nu \nabla^2 \vec{u} + \lambda (\vec{u}^s - \vec{u}) \chi,
\end{align}
where $\vec{u} = \vec{u}(\vec{x}, t)$ is the fluid velocity field, $\rho = 1$ the fluid density, $p = p(\vec{x}, t)$ the pressure, $\nu = 10^{-4}$ the kinematic viscosity, $\lambda = 10^6$ Brinkman penalization coefficient, $\vec{u}^s = \vec{u}^s(t)$ the velocity of the cylinder and $\chi = \chi(\vec{x}, t)$ the characteristic function of the cylinder, equal to $1$ inside the cylinder and $0$ outside it.
The equation is solved on a $[0, 1] \times [0, 0.5]$ domain with open boundary conditions.
A solid cylinder of diameter $d = 0.075$ is fixed at the coordinate $(0.2, 0.25)$ relative to the simulation domain.
The cylinder and the simulation domain are moving horizontally at the speed of $u^s_x(t) = \Re(t) \nu / d$ relative to the fluid, with Reynolds number $\Re(t)$ (the external forcing) varying between $\Re = 400$ and $\Re = 1200$.
In this range, for a fixed $\Re$, a vortex street forms behind the cylinder.

The~\cref{eq:ns-incompressible,eq:ns-brinkman} are solved using a pressure projection method~\cite{chorin1968numerical} on an adaptive Cartesian mesh of maximum resolution of $1024 \times 512$ cells.
For the purpose of this study, the adaptive, non-uniform mesh is interpolated to the maximum resolution of $1024 \times 512$ cells and is thus treated as a uniform mesh when used by AdaLED.

\begin{figure}
  \centering
  \includegraphics[width=\columnwidth]{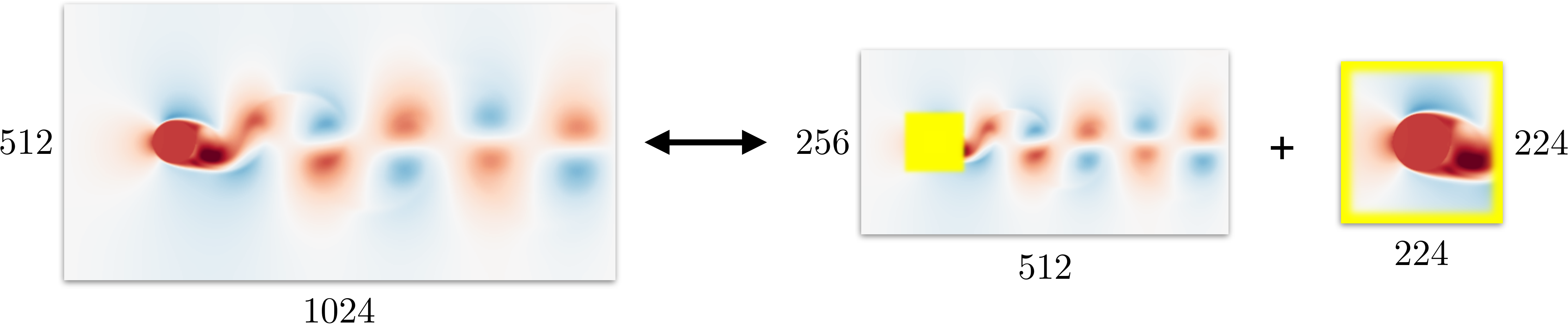}
  \caption{
    Schematic view of the multiresolution AE.
    To accelerate the training and reduce the storage and memory requirements, the AE operates on two downsampled variants of the velocity field.
    The yellow region denotes the blending mask used for reconstructing the full resolution field.
  }
  \label{fig:mr-compression}
\end{figure}

In an effort to reduce the computational demands of the high-dimensional grid and speed up the training process of AdaLED, we propose a novel multiresolution physics-based AE.
The proposed AE takes advantage of the characteristics of the flow and uses a reduced-resolution grid far to the cylinder where the flow exhibits simpler features compared to the vicinity of the cylinder.
Concretely, the AE operates on two downsampled grids: a half-resolution grid spanning the entire domain and a small full-resolution patch around the cylinder (\cref{fig:mr-compression}).
By utilizing this multiresolution approach, we are able to reduce the storage and memory requirements and speed up the training.
Furthermore, the AE outputs the stream function instead of the velocity field~\cite{mohan2020embedding}.
This physics-inspired architecture ensures zero divergence of the velocity field (as per \cref{eq:ns-incompressible}).
An additional physics-based vorticity loss is added to improve performance.
The specifics are outlined in \cref{sec:cfd-autoencoder}.

In addition to the latent state $\vz(t)$, which is necessary to recreate the system dynamics and support macro-to-micro transitions, the MLM also outputs the force $\vFcyl(t)$ exerted by the fluid on the cylinder as the quantity of interest $\vq(t)$.
The relative importance of $\vFcyl$ and $\vz$ in the macro propagator's training loss and uncertainty estimation can be controlled by scaling the force with some factor $\alpha_F$.

The macro propagator is an ensemble of five probabilistic LSTMs.
Each LSTM is trained to predict the mean and the variance of the residuals $\Delta \vz(t) = \vz(t + \Delta t) - \vz(t)$ and $\Delta \vq(t) = \vq(t + \Delta t) - \vq(t)$, $\vq(t) = \alpha_F \vFcyl(t)$.
The acceptance criterion is based on a relative reconstruction error $E$ of the velocity field
\begin{equation}
  \label{eq:cup-E}
  E(\tilde{\vu}, \vu) = \frac{\norm{\vu - \tilde{\vu}}_2^2}{\norm{\vu}_2^2}, \quad
  \norm{\vu}_2^2 = \sum_{ij} \vu_{ij}^2,
\end{equation}
where $\vu$ is the full-resolution velocity field from the micro propagator, and $\tilde{\vu} = \Decoder^{\vtheta_\Decoder}(\tilde{\vz})$ the prediction of the ML model.
The autoencoder itself is trained on a different loss function, explained in \cref{sec:cfd-autoencoder}.
The total uncertainty $\sigma$ of the prediction of the macro propagator is defined as the standard deviation of the uncertainty vectors $\sigma = \sqrt{ \left( (\vsigma_{\Delta \vz}(t))^2 + (\vsigma_{\Delta \vq})(t)^2 \right) / (d_z + 2) }.$

The time step of AdaLED is set to $\Delta t = 0.005$, resulting in approximately $60$ AdaLED time steps per vortex street period for $\Re = 1000$.
The internal time step of the micro propagator $\delta t$ is, for simplicity, fixed throughout the simulation.
For simulations with $\Re(t)$ of up to 1000, $\delta t = 0.005/18$, and for simulations with $\Re(t)$ of up to 1200, $\delta t = 0.005/21$, resulting in a Courant number of ${\sim}0.4$.

We use AdaLED cycles of 4 warm-up steps, 12 online validation steps, between 400 and 500 macro steps, and between 9 and 14 micro steps.
Both limits are chosen uniformly at random for each cycle to avoid synchronizing AdaLED cycles with vortex street periods.
The capacity of the dataset is set to 256 trajectories.
To maximize the speed-up of AdaLED, training is performed on a separate compute node in parallel with the inference and the micro propagator, as depicted in \cref{fig:server-client}.
Experiments were conducted on the Piz Daint supercomputer on two XC50 nodes, each equipped with one 12-core Intel Xeon E5-2690 CPU running at 2.6 GHz and one Nvidia P100 16GB GPU.
The simulations were performed using the \cubismAMR{} software~\cite{chatzimanolakis2022cubismamr}.

\subsection{Results}

In~\cref{sec:cup-selected}, we demonstrate the effectiveness of AdaLED in accelerating the simulation of the flow past the cylinder without sacrificing accuracy.
In~\cref{sec:cup-training-study}, we highlight the importance of adaptive training in systems with changing dynamics.
Finally, in \cref{sec:mr-study}, we conduct an ablation study to evaluate the advantage of the multiresolution autoencoder.

\subsubsection{Effectiveness of AdaLED}
\label{sec:cup-selected}

We perform the simulation for a total of \num{300000} time steps, with Reynolds number transitioning cyclically between $\Re = 600$, $\Re = 750$, and $\Re = 900$ every \num{5000} time steps.
The hyper-parameters, listed in \cref{tbl:mr-search-space}, are tuned according to the performance on a shorter simulation, as reported in \cref{sec:mr-study}.
For this simulation, the error and uncertainty thresholds are set to $E_\text{max} = 0.017$ and $\sigma^2_\text{max} = 0.00035$, respectively.
We note that these are the key AdaLED hyper-parameters that can be adjusted to balance accuracy and acceleration as desired.

\begin{figure}
  \centering
  \includegraphics[width=\columnwidth]{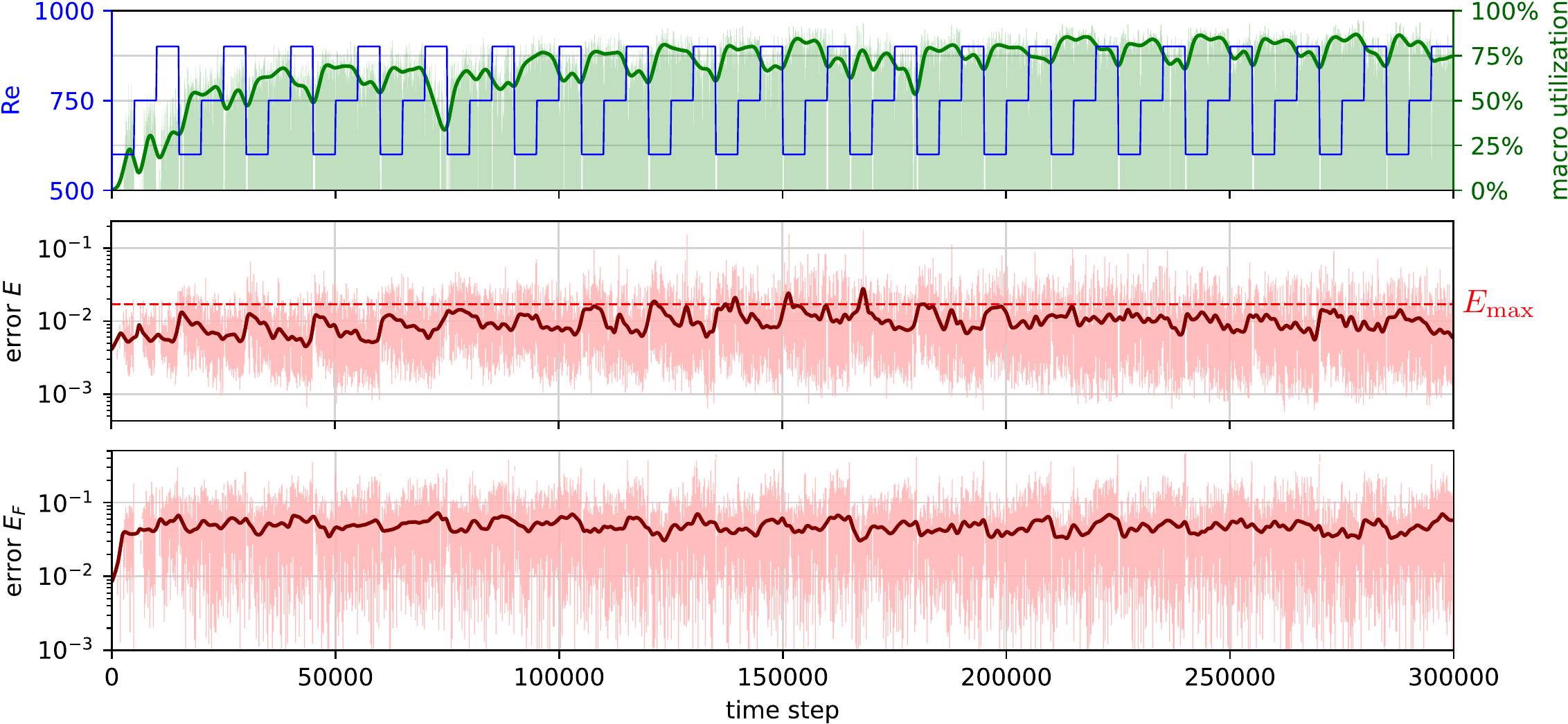}
  \caption{
    AdaLED performance on a flow behind cylinder simulation for $\Re(t) \in \{600, 750, 900\}$ (\cref{sec:cup-selected}).
    Top: Reynolds number $\Re(t)$ profile and the macro utilization $\eta$.
    Middle and bottom: validation errors of the velocity ($E$, \cref{eq:cup-E}) and force on the cylinder ($E_F$, \cref{eq:cup-EF}).
    The per-step errors (faded red) alternate between low values at the beginning of the macro-only stage and higher errors at the end of the macro-only stage.
    The errors for velocity stay close to $1\%$ on average (dark red) and close to $5\%$ for the force (with a cross-correlation of 0.99).
    The errors refer only to macro-only steps.
    A detailed view of errors in a short simulation section is shown in \cref{fig:selected-validation-vF}.
  }
  \label{fig:selected-utilization}
\end{figure}

The $\Re(t)$ profile, the macro utilization, and the errors are displayed in \cref{fig:selected-utilization}.
The errors correspond to validation errors during the online validation phase and test errors during the macro-only stage.
To calculate these errors, we compare the micro states $\vu_t$ with the reconstructed states $\tilde{\vu}_t = \Decoder^{\vtheta_\Decoder}(\tilde{\vz}_t)$ produced by the MLM.
We use the velocity field error metric $E$ from \cref{eq:cup-E} to quantify the error in the velocity field.
For the force $\vFcyl$ error, we define a normalized error~$E_F$ as follows:
\begin{equation}
  \label{eq:cup-EF}
  E_F = E_F(\vFcyl', \vFcyl) = \frac{\norm{\vFcyl' - \vFcyl}}{F_\text{cyl}^\text{avg}},
\end{equation}
where $F_\text{cyl}^\text{avg} = \left< \norm{\vFcyl} \right> \approx 0.079$ is the average magnitude of the force.
For testing purposes, to retrieve the micro states $\vu_t$, we continue running the simulation even in the macro-only stages.
The training of the MLM is temporarily suspended during this time.

In \cref{fig:selected-utilization}, we observe the same trend as in the previous two case studies.
Initially, the macro utilization is zero.
After the networks become sufficiently trained, the framework starts to accept the predictions of the MLM.
As training continues, the errors and uncertainties decrease, resulting in an increase in macro utilization.
During the macro-only stage, the testing error $E$ and $E_F$ remain low, averaging to $1\%$ and $5\%$, respectively.
The errors can be further decreased at the cost of reduced speed-up.

\begin{figure}
  \centering
  \includegraphics[width=\columnwidth]{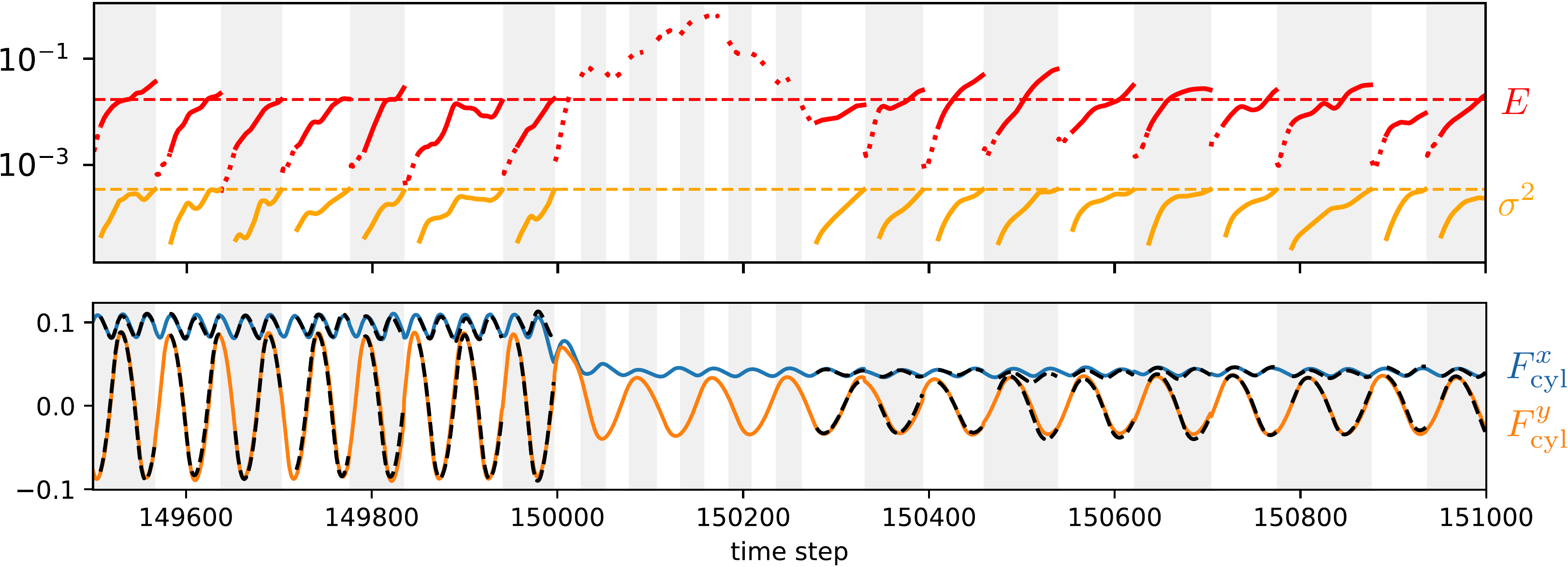}
  \caption{
    A detailed view of the part of the simulation of flow behind the cylinder from \cref{sec:cup-selected} and \cref{fig:selected-utilization}, during the $\Re = 900$ to $\Re = 600$ transition.
    Top: velocity validation error $E$ and the squared uncertainty $\sigma^2$  (dotted for warm-up and online validation stages, solid for macro-only) and their thresholds $E_\text{max}$ and $\sigma^2_\text{max}$ (dashed).
    Bottom: horizontal force on the cylinder (blue), vertical force (orange), and the macro's prediction (dashed black).
  }
  \label{fig:selected-validation-vF}
\end{figure}

In \cref{fig:selected-validation-vF}, we present a closer look at how error and uncertainty change over a selected section of the trajectory, specifically during the transition from $\Re = 900$ to $\Re = 600$.
The acceptance of the macro prediction is determined by the error $E$ and its threshold $E_\text{max}$.
Before the $\Re$ transition, the online validation error remains below the threshold, and AdaLED accepts the macro prediction.
However, during the transition period, which lasts for a few hundred time steps, the MLM cannot reliably predict the dynamics.
Hence, AdaLED switches to the micro propagator instead.
Once the transition period ends, AdaLED resumes utilizing the macro propagator.
It should be noted that the error $E$ may exceed the threshold $E_\text{max}$ at times.
The threshold $E_\text{max}$ represents the maximum error at the end of the online validation stage, so it should be set to a value lower than the desired maximum tolerable error.

In contrast to the error $E$ that controls whether AdaLED enters the macro-only stage, the uncertainty $\sigma$ and the threshold $\sigma_\text{max}$ control its duration.
Once $\sigma$ exceeds $\sigma_\text{max}$, the macro-only stage is stopped.

The accuracy of the predicted force $\vFcyl$ is illustrated in the bottom plot of~\cref{fig:selected-validation-vF}.
We observe a good agreement between the two force profiles.
A visual representation of the latent trajectory can be found in \cref{fig:selected-latent-F} in \cref{app:cyl-latent}.

\begin{figure}
  \centering
  \includegraphics[width=\columnwidth]{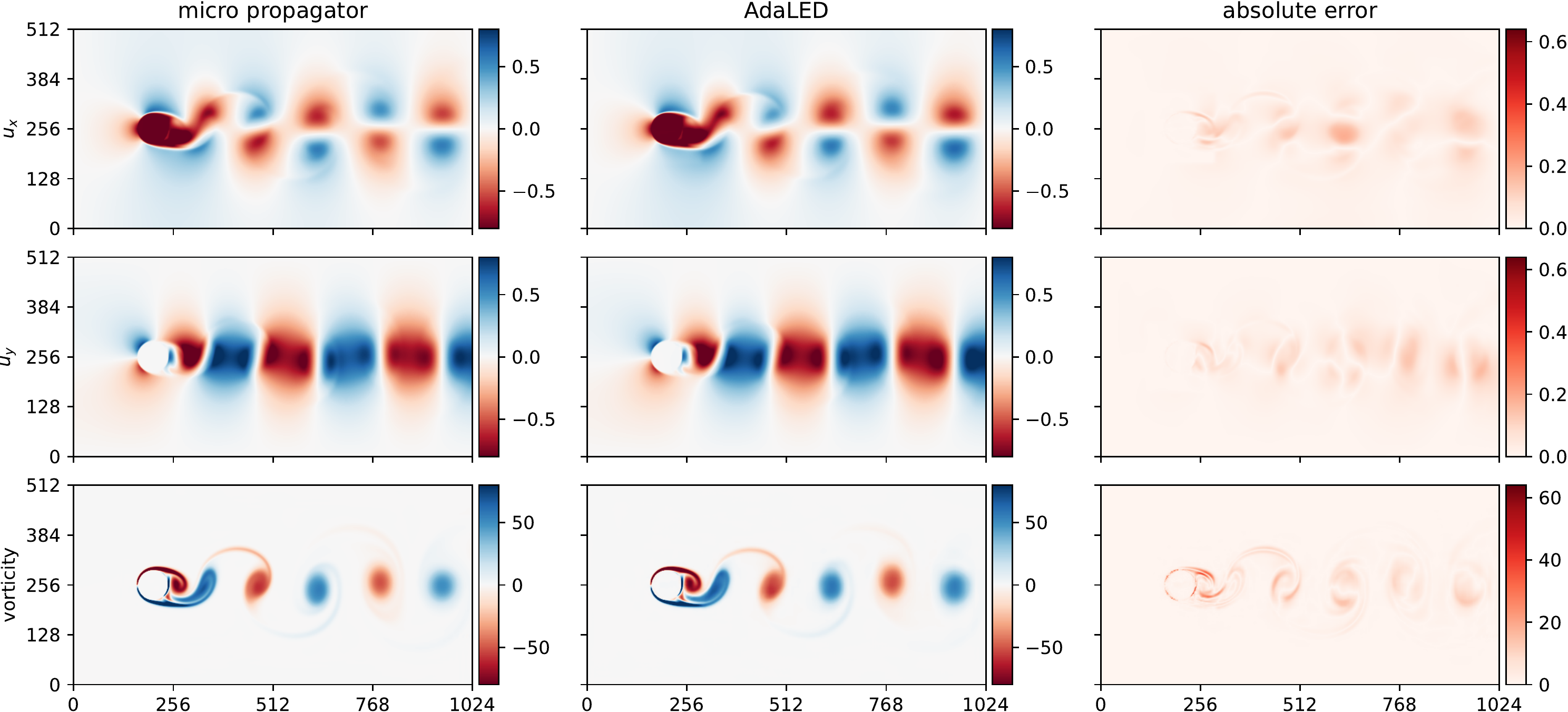}
  \caption{
    Snapshot of the time step 212500 ($\Re = 600$) of the simulation from \cref{sec:cup-selected}, 50 time steps into the macro-only stage, with a relative error of $E \approx 0.014$ (\cref{eq:cup-E}).
    Left: micro propagator state $\vu_t$ and vorticity $\omega_t$, middle: the prediction of the surrogate (MLM) and full-resolution reconstruction, right: absolute error.
  }
  \label{fig:selected-reconstruction}
\end{figure}

A snapshot of the simulation during the macro-only stage is visualized in \cref{fig:selected-reconstruction}.
We observe that AdaLED reproduces the state of the simulation accurately and captures the characteristics of the flow with high accuracy.
Notably, errors concentrate on the fine-scale structures of the flow.
Arguably, the double arcs in the error profile indicate that the error can be partially attributed to the macro propagator advancing the dynamics at an incorrect speed.

\begin{figure}
  \centering
  \includegraphics[width=0.95\columnwidth]{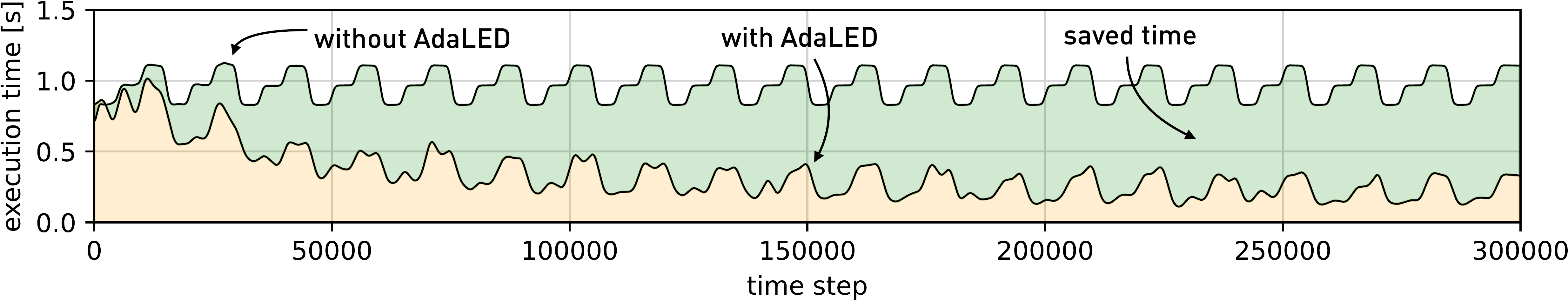}
  \caption{
    Smoothed time step execution time with and without AdaLED for the simulation setup from \cref{sec:cup-selected}.
    The speed-up factor converges to 4.3x (\cref{tbl:cup-execution-time}).
    The periodic changes in the execution time correspond to the periodic changes of the Reynolds number (see the top plot in \cref{fig:selected-utilization}).
  }
  \label{fig:cup-execution-time}
\end{figure}

The execution time of the standalone simulation and the AdaLED-accelerated simulation is compared in \cref{fig:cup-execution-time}.
The green area represents the time saved by using AdaLED.
The total macro utilization over the whole run (\num{300000} time steps) is $69\%$, achieving a speed-up of approximately 2.9x.
After the training converges, the macro utilization reaches $80\%$ in the last \num{15000} time steps, resulting in a speed-up of 4.3x.
This implies that the trained MLM can be applied to other simulations with $\Re \in \{600,750,900\}$ achieving similar performance.

The execution time breakdown of each simulation time step is presented in \cref{tbl:cup-execution-time}.
By itself, the micro propagator takes on average \SI{969}{ms} per time step for the given profile of $\Re(t)$ (larger $\Re$ are slower to simulate).
When used within AdaLED, after the first micro-to-macro transition happens, the imperfect autoencoder reconstruction causes a slight increase in the mesh size and thus slows the micro propagator down to \SI{1055}{ms} per step (+9\%).
The average overhead of AdaLED (autoencoders, logging, diagnostics, etc.) is relatively small, averaging to \SI{42}{ms} per time step on average.
Finally, the macro propagator requires only \SI{5}{ms} per time step.

We conclude that for computationally expensive CFD simulations, the overhead for deploying AdaLED is minimal.
In this case, the speed-up is determined effectively only by the macro utilization.
Higher speed-ups can be achieved by affording higher errors (increasing the error thresholds) or employing models with higher accuracy.
The latter can be achieved through network architecture improvements, more effective training procedures, larger ensembles, or more extensive tuning.

The results of a run with a different Reynolds number profile are shown in \cref{app:cyl-Re-generalization}.

\begin{table}
  \caption{
    Average execution times and their standard deviation for different time step components for the simulation setup from~\cref{sec:cup-selected}.
  }
  \centering
  \begin{tabular}{l|c}
    step or partial step & execution time {[ms]} \\
    \hline
    Without AdaLED & \\
    \quad \textbf{micro propagator} & $\mathbf{969 \pm 104}$ \\
    With AdaLED & \\
    \quad micro propagator & $1055 \pm 113$ \\
    \quad overhead of AdaLED & $42 \pm 1$ \\
    \quad macro-only step & $4.6 \pm 0.3$ \\
    \quad \textbf{average} & $\mathbf{339 \pm 183}$ \textbf{(2.9x speed-up)} \\
    \quad \textbf{average} (last 15k time steps) & $\mathbf{225 \pm 79}$ \textbf{(4.3x speed-up)} \\
  \end{tabular}
  \label{tbl:cup-execution-time}
\end{table}

\subsubsection{The importance of adaptivity}
\label{sec:cup-training-study}

{
\definecolor{A}{RGB}{1,115,178}
\definecolor{B}{RGB}{222,143,5}
\definecolor{C}{RGB}{2,158,115}
\definecolor{D}{RGB}{213,94,0}
\newcommand{\SA}{{\color{A} A}}
\newcommand{\SB}{{\color{B} B}}
\newcommand{\SC}{{\color{C} C}}
\newcommand{\SD}{{\color{D} D}}

In the following, we demonstrate the importance of adaptivity, i.e., constantly training throughout the whole simulation and adapting to new states and trajectories, compared to pretraining or training only until a given point in time.
We analyze two profiles of time-varying Reynolds numbers $\Re(t)$.
In the first, $\Re(t)$ switches between values $500$, $750$, and $1000$ in a zig-zag fashion throughout the whole simulation.
In the second, $\Re(t)$ starts as the first profile but switches to a different regime (400, 600, 800, 1000, and 1200) in the second half of the simulation, to emulate a system that enters a new regime late in the simulation.
For each profile, two setups are tested: one with training enabled all the time (\emph{adaptive}) and one with training enabled only at the first half of the simulation (\emph{non-adaptive}).

Apart from testing adaptivity, we test how disabling micro-to-macro transitions in the first half of the simulation affects the quality of the MLM in the second half.
Namely, we expect that delaying initial transitions and providing more time for training may help improve the accuracy and macro utilization in the later stages of the simulation.
Thus, for each $\Re(t)$ profile, we test in total four setups: adaptive without delay (A; default AdaLED behavior), non-adaptive without delay (B), adaptive with delay (C), and non-adaptive with delay (D).
For each setup, five runs with different random seeds are performed to obtain the variance in performance.

The macro utilization and relative MSE on the velocity for the first $\Re(t)$ profile are visualized in \cref{fig:cup-training-same-Re}.
We observe that training only in the first half with transitions disabled (the setup D) achieves higher accuracy in the second half of the simulation compared to other setups.
In fact, the non-adaptive setup D exhibits higher macro utilization and lower error compared to the adaptive setup C.
This is expected as the training dataset from the first half of the run already contains all the information needed to forecast effectively the dynamics in the second half (the profiles are similar).
As a consequence, there is no need for online training.

However, we observe a different phenomenon when the system regime changes over time, as shown in \cref{fig:cup-training-different-Re}.
Here, the macro utilization in non-adaptive setups B and D drops to zero when the system enters the previously unseen $\Re(t) = 1200$ regime, whereas the adaptive setups A and C eventually adapt and achieve macro utilization of $25$ to $35\%$.

We note here that the available training time for setups D and C is higher.
While setups A and B accelerated the simulation from the start and had only 6h for training during the first half of the simulation, setups D and C took 11h for the first half and thus had almost twice as much time for training before being tested in the second half.
We argue that this phenomenon is an important characteristic property of online surrogates, i.e., the higher speed-up they achieve, the less time they have for training.

\DeclareRobustCommand\full   {{\color{A} \tikz[baseline=-0.6ex]\draw[very thick] (0,0)--(0.5,0);}}
\DeclareRobustCommand\dotted {{\color{B} \tikz[baseline=-0.6ex]\draw[very thick,dotted] (0,0)--(0.54,0);}}
\DeclareRobustCommand\dashed {{\color{C} \tikz[baseline=-0.6ex]\draw[very thick,dashed] (0,0)--(0.54,0);}}
\DeclareRobustCommand\dashdot{{\color{D} \tikz[baseline=-0.6ex]\draw[very thick,dash dot] (0,0)--(0.5,0);}}
\begin{figure}
  \centering
  \includegraphics[width=\textwidth]{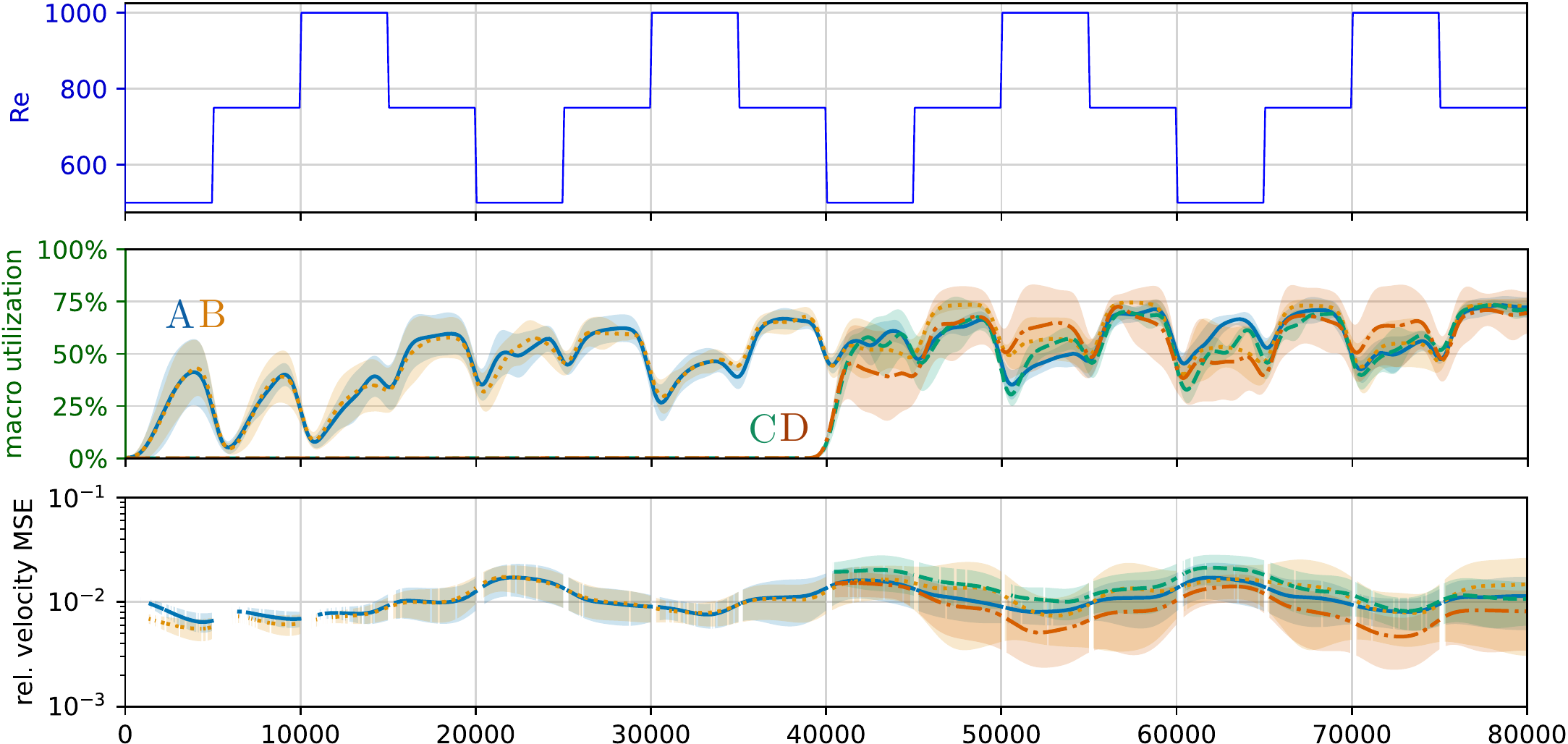}
  \caption{
    The adaptivity and transition delay study from \cref{sec:cup-training-study} for the $\Re(t)$ with a fully repeating profile.
    Lines denote:
    adaptive without delay (\SA~\full),
    non-adaptive without delay (\SB~\dotted),
    adaptive with delay (\SC~\dashed),
    non-adaptive with delay (\SD~\dashdot).
    Top: Reynolds number profile, middle: macro utilization $\eta$, bottom: smoothed relative MSE of the velocity in macro-only stages.
    Shaded regions, where available, denote the standard deviation along five repeated runs of the same setup.
  }
  \label{fig:cup-training-same-Re}
\end{figure}

\begin{figure}
  \centering
  \includegraphics[width=\textwidth]{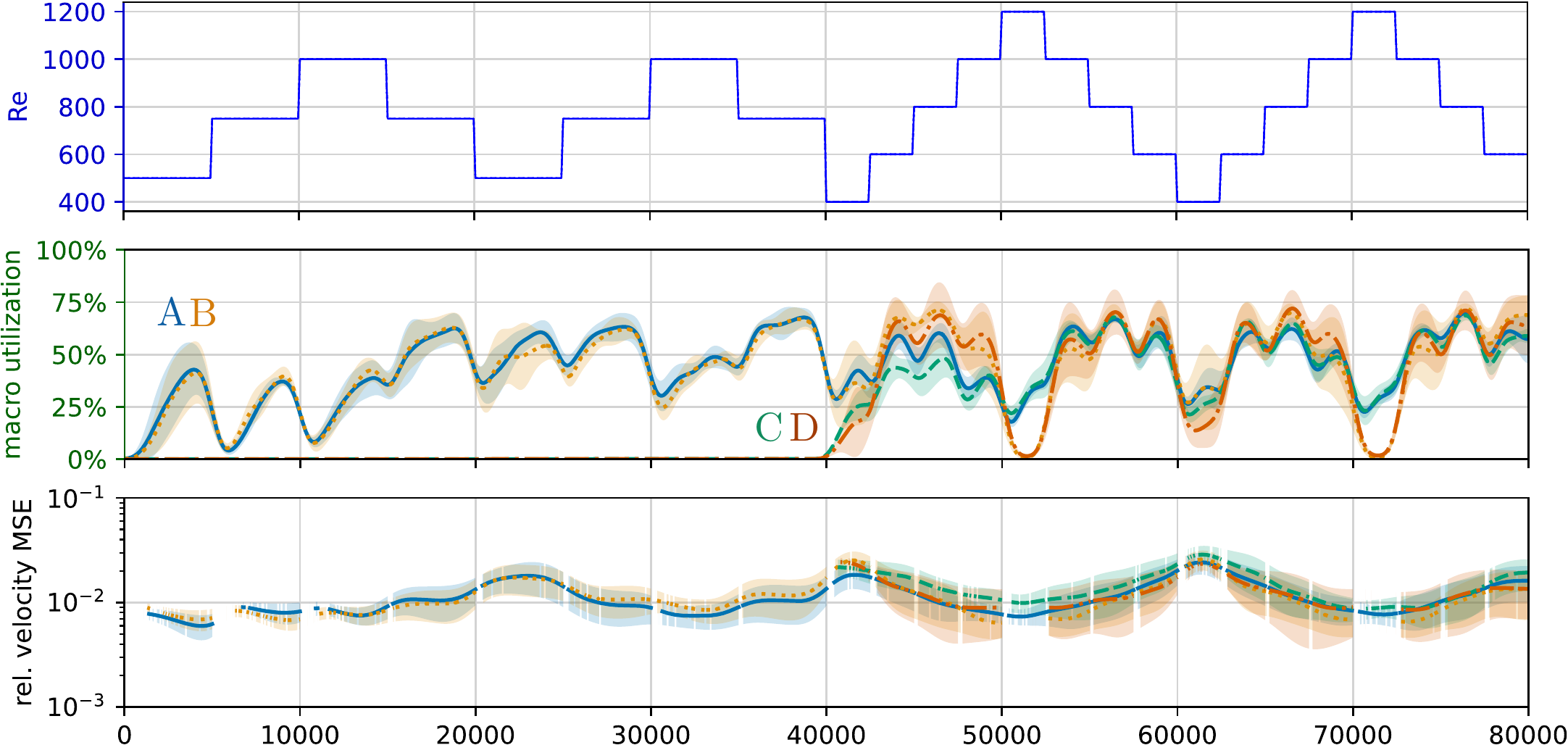}
  \caption{
    Analogous of \cref{fig:cup-training-same-Re}, for the $\Re(t)$ that changes the profile in the second half of the simulation.
    The drop in performance in the second half is clearly visible for the non-adaptive cases (\SB~\dotted and \SD~\dashdot).
  }
  \label{fig:cup-training-different-Re}
\end{figure}
}

\subsubsection{Multiresolution autoencoders}
\label{sec:mr-study}

In this section, we perform an ablation study to analyze the benefit of the multiresolution convolutional autoencoder.
We compare three cases: (i) single resolution, (ii) multiresolution with a $256 \times 256$ patch around the cylinder, and (iii) multiresolution with a $224 \times 224$ patch.
For each case, we perform 80 runs with randomized thresholds $E_\text{max}$ and $\sigma_\text{max}$, learning rates, force scaling $\alpha_F$, latent state size $d_z$, number of CNN channels, and the number of layers.
The hyper-parameter search space is listed in \cref{tbl:mr-search-space}.
The CNN architecture, the remainder of hyper-parameters, and the breakdown of training execution time are described in~\cref{sec:cup-hyperparam-and-cnn}.
The simulation setup matches the one from~\cref{sec:cup-selected} (Reynolds number $\Re(t)$ cycles between 600, 750, and 900 every \num{5000} time steps), with a shorter running time of \num{60000} time steps.
On average, a single simulation run takes approximately 16 hours to complete.
Three performance metrics are considered: total macro utilization $\eta$, average velocity relative MSE $E$ (\cref{eq:cup-E}), and the average force error $E_F$ (\cref{eq:cup-EF}).
The averages include only the macro-only stages.

\begin{figure}
  \centering
  \includegraphics[width=0.7\columnwidth]{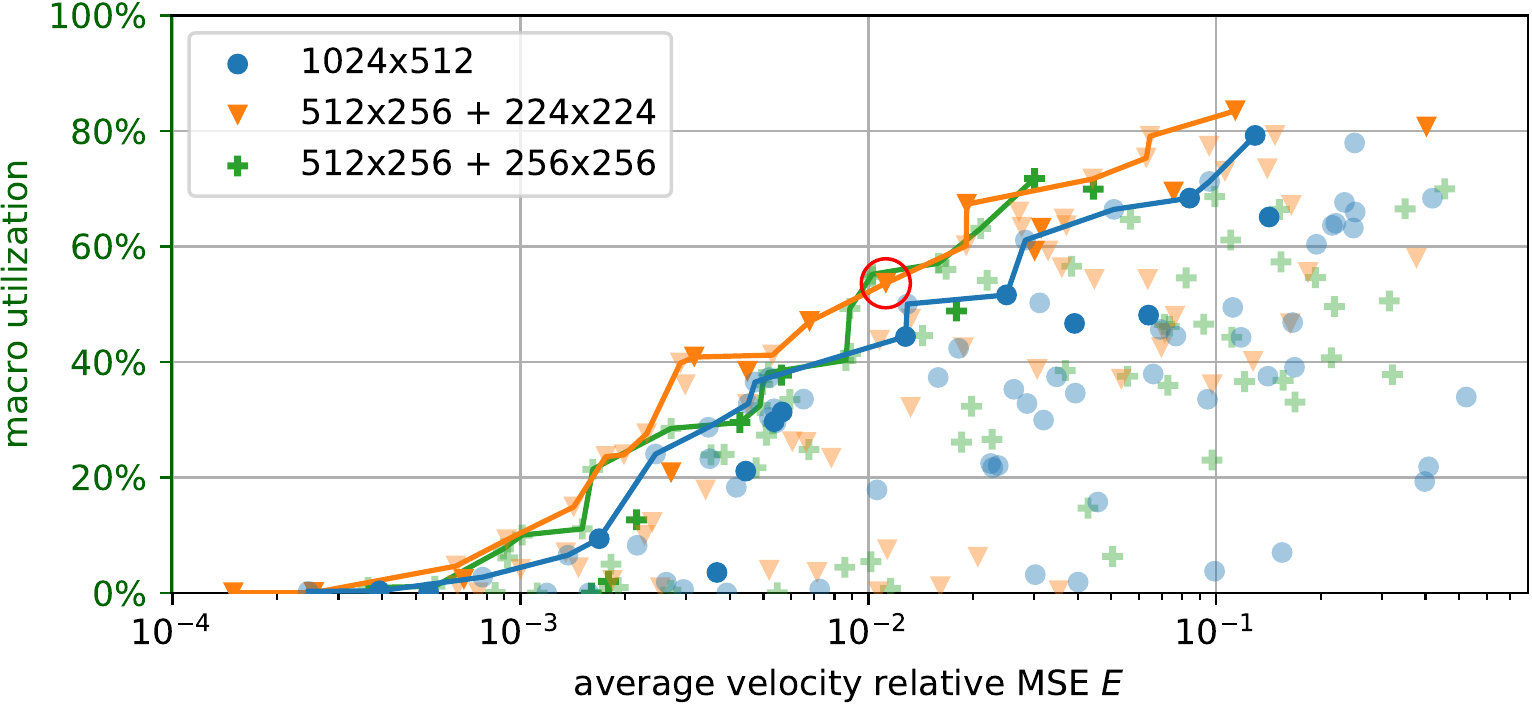}
  \caption{
    Multiresolution autoencoder multi-objective study (\cref{sec:mr-study}), optimizing for total macro utilization $\eta$ and velocity error $E$.
    The lines represent the Pareto fronts for each autoencoder setup.
    Darker symbols denote samples that are also optimal in the $\eta$--$E_F$ sense (\cref{fig:mr-pareto-F}).
    The encircled sample is the reference parameter set used in the rest of the study.
  }
  \label{fig:mr-pareto-v}
\end{figure}

\begin{figure}
  \centering
  \includegraphics[width=0.7\columnwidth]{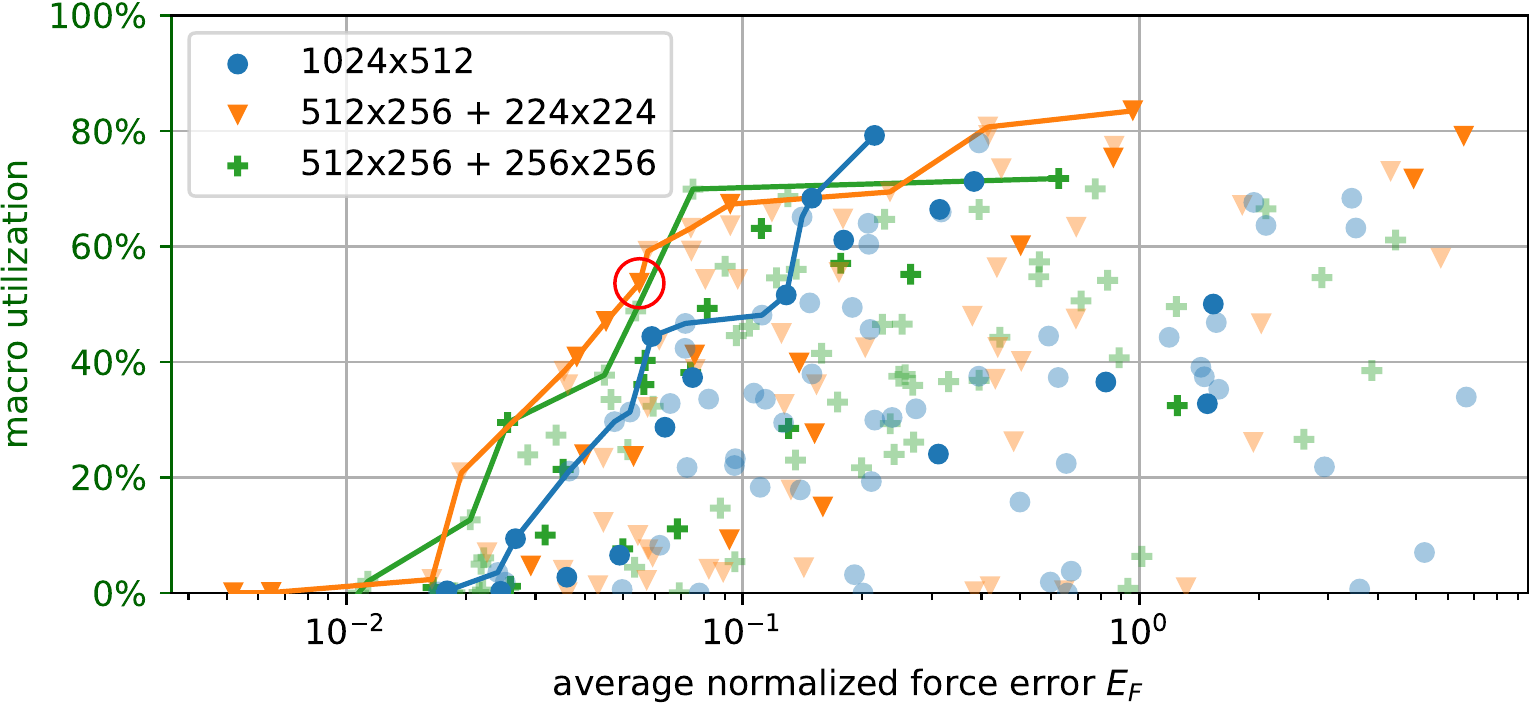}
  \caption{
    Analogous of \cref{fig:mr-pareto-v}, with $x$-axis denoting the average normalized cylinder force root MSE $E_F$ instead of the velocity field error $E$.
    Here, darker symbols denote optimal samples in the $\eta$--$E$ sense.
  }
  \label{fig:mr-pareto-F}
\end{figure}

The results of the comparison are shown in \cref{fig:mr-pareto-v,fig:mr-pareto-F}, where the macro utilization is plotted against the average errors $E$ and $E_F$, respectively.
We observe a clear advantage of the multiresolution approach compared to the single-resolution autoencoder in terms of both macro utilization and accuracy.
The encircled point in the plots refers to the hyper-parameter set used in \cref{sec:cup-selected,sec:cup-training-study}, which resulted in a macro utilization of $54\%$ in \num{60000} time steps ($69\%$ when run for \num{300000} time steps).
The performance of this hyper-parameter set on a different Reynolds number profile is shown in \cref{app:cyl-Re-generalization}.
It is important to note that the results are subject to random variations, with errors varying by about $\pm 5\%$ (relative) and macro utilization varying by $\pm 3\%$ (absolute), depending on the random seed.

\begin{table}
  \caption{
    Parameter search space for the multiresolution autoencoder study, and the selected parameter set.
  }
  \centering
  \begin{tabular}{c|cc|c}
    parameter & \multicolumn{2}{c|}{search space} & selected \\
    \hline
    multiresolution?       & no & yes & yes \\
    inner resolution       & N/A & \{256x256, 224x224\} & 224x224 \\
    \# of CNN layers       & \{5, 6\} & \{4, 5\} & 4 \\
    channels/layer         & \multicolumn{2}{c|}{\{16, 20, 24\}} & 16 \\
    $d_z$ (per resolution) & \multicolumn{2}{c|}{\{4, 8, 12, 16, 24\}} & 8 \\
    AE learning rate       & \multicolumn{2}{c|}{$\LogUniform(0.0001, 0.001)$} & 0.00047 \\
    LSTM learning rate     & \multicolumn{2}{c|}{$\LogUniform(0.0003, 0.003)$} & 0.00126 \\
    scaling $\alpha_F$     & \multicolumn{2}{c|}{$\LogUniform(0.03, 30.0)$} & 7.2 \\
    $E_\text{max}$         & \multicolumn{2}{c|}{$\LogUniform(0.001, 0.1)$} & 0.017 \\
    $\sigma^2_\text{max}$    & \multicolumn{2}{c|}{$\LogUniform(0.00001, 0.1)$} & 0.00035 \\
  \end{tabular}
  \label{tbl:mr-search-space}
\end{table}

\section{Discussion}
\label{sec:discussion}

We present AdaLED, a framework that employs CAEs and an ensemble of RNN-LSTMs to learn data-driven, online, adaptive machine-learned models to accelerate the simulations of complex systems.
The model is trained in parallel with the original simulation (micro propagator) and can adapt online to newly discovered dynamics.
More importantly, the model monitors its accuracy and prediction uncertainty and replaces the micro propagator only when its accuracy is high, and its uncertainty is low.
This mechanism enables the acceleration of simulations for sections of the state space that are learned, even if the model cannot or is not yet fully trained to faithfully reproduce the whole complex state space dynamics.
In regions of the state that are underrepresented or not part of the training data, AdaLED utilizes the original computational solver.

We demonstrate AdaLED in three benchmark problems: the Van der Pol oscillator dynamics with varying parametric nonlinearity $\mu \in [1,3]$, a 2D reaction-diffusion equation with varying diffusion parameter $d \in [0.1, 0.2]$, and flows past a circular cylinder at varying $\Re \in [400, 1200]$. 
On these benchmarks, we demonstrate its ability to train a machine-learned model progressively, exploit its predictions only when they are reliable, and detect when a system enters an unseen regime in the phase space.
The trained model demonstrates high accuracy in all dynamic regimes seen during training and does not suffer from catastrophic forgetting.

On the flow past a cylinder at varying $\Re$, AdaLED, starting from untrained networks, reproduces the dynamics of vastly different dynamical regimes, achieving a net speed-up of ~2.9x for a 3-day-long simulation.
This speed-up is achieved at the cost of a mean square error of only ${\sim}1\%$ on the velocity field, ${\sim}5\%$ root mean square error of the force on the cylinder and cross-correlation of 0.99.
The speed-up can be increased further at the cost of lower accuracy by increasing $E_\text{max}$ and $\sigma_\text{max}$.
We emphasize the advantage of AdaLED compared to other frameworks to control this trade-off between speed-up and accuracy.
To our knowledge, AdaLED is the first method that can efficiently learn to propagate the high-dimensional dynamics of a complex flow at various regimes using a single surrogate, offering a robust accuracy vs. speed-up trade-off.

Our findings suggest that AdaLED is a potent adaptive algorithm for adaptively constructig and interfacing surrogates that  acceleratecomplex multiscale simulations. 
We believe that AdaLED can be employed as a black box accelerator that takes advantage of repeating patterns in computation-heavy tasks.
In the future, we plan to investigate its acceleration capabilities on reinforcement learning tasks and model parameter optimizations, where multiple simulations can share the same surrogate.

Moreover, we argue that the proposed framework is a valuable contribution to the digital twin literature~\cite{rasheed2020digital,kapteyn2021probabilistic}.
AdaLED combines data assimilation, real-time monitoring, and online adaptive data-driven learning to build a surrogate.
The proposed framework is directly applicable if the simulation of the physical system is possible from any initial condition at will.
Otherwise, it can be applied only with minor modifications (by turning off the restarting of the micro-scale solver).
This way, the framework can be employed to learn a digital replica of a physical system, i.e., the digital twin. 
The surrogate's response under different conditions and parametrizations can be tested at will, avoiding the cost and computational burden of experiments or fully resolved simulations and the risk of exposing the original system to adverse conditions~\cite{vinuesa2022enhancing,vinuesa2023transformative}.

Application-wise, AdaLED can benefit from more advanced autoencoders, such as variational autoencoders~\cite{kingma2013auto}, autoencoders that take into account temporal correlations~\cite{girin2020dynamical}, or non-uniform autoencoders based on space-filling curves~\cite{heaney2020applying} and octrees~\cite{riegler2017octnet,tatarchenko2017octree}.
A topic of ongoing research is to utilize the latter to help scale AdaLED to 3D fluid flows.
Moreover, all latent state variables are currently treated as equally important.
The method could benefit from compression techniques that can estimate the relevance of each latent dimension in the reconstruction~\cite{fukami2020convolutional}.

Recently proposed hierarchical deep learning time-steppers~\cite{liu2022hierarchical}, reduced-order propagators on the latent space~\cite{pawar2019deep}, and Autoformer networks~\cite{wu2021autoformer} demonstrate promising results in PDEs and other complex time-series data.
These algorithms can be employed as efficient macro propagators.
Having a very fast macro propagator opens space for more advanced techniques, such as planning optimal actions in reinforcement learning~\cite{moerland2020model}.

Likewise, if the application allows it, we could detect dynamic regimes underrepresented in the data by simulating many steps in advance and looking at the future prediction uncertainty to determine if we should perform a macro-to-micro transition early.
An additional network could be trained to estimate the decoder reconstruction error given the current latent state.
Combined with the macro propagator, the macro-to-micro transition criteria could be based on the joint uncertainty of the ensemble and this reconstruction error.

Finally, we plan to investigate improved scheduling and refined control of AdaLED cycles and the micro-macro transitions to reduce the total number of time steps performed in the micro-scale and to provide more control over the trade-off between the adaptivity versus speed-up.
In this direction, AdaLED can benefit from novel algorithms for uncertainty quantification of supervised learning algorithms~\cite{egele2022autodeuq}.

\section{Acknowledgements}
\label{sec:acknowledge}
We acknowledge support from The European High-Performance Computing Joint Undertaking (EuroHPC) Grant DCoMEX (956201-H2020-JTI-EuroHPC-2019-1),
and computing resources from the Swiss National Supercomputing Centre (CSCS) under projects s930 and s1160.
We would like to thank Pascal Weber (ETHZ) for several useful discussions.

\section{Data and Code Availability}
\label{sec:code}
The source code will be made readily available at \url{https://github.com/cselab/adaled} upon publication.

\bibliographystyle{elsarticle-num}
\bibliography{bibliography}

\counterwithin*{figure}{section}
\counterwithin*{table}{section}

\appendix
\section{Macro propagator LSTMs}
\label{sec:lstm}

{  %
\renewcommand{\vb}{\vec{b}}
\newcommand{\vxi}{\bm{\xi}}
\newcommand{\vc}{\vec{c}}
\newcommand{\vf}{\vec{f}}
\newcommand{\vg}{\vec{g}}
\newcommand{\vi}{\vec{i}}
\newcommand{\vo}{\vec{o}}
\renewcommand{\W}{\vec{W}}
\newcommand{\wR}{\vec{\omega}_R}
\newcommand{\wH}{\vec{\omega}_H}

The macro propagator of AdaLED is an ensemble of multi-layer long short-term memory (LSTM)~\cite{hochreiter1997long} recurrent neural networks (RNNs).
Given an input state $\vxi_t \in \RR^{d_\xi}$, the current hidden state $\vh_t \in \RR^{d_h}$ and the cell state $\vc_t \in \RR^{d_h}$, each layer of each network computes the next hidden state $\vh_{t+\Delta t}$ and the cell state $\vc_{t+\Delta t}$ as follows (layer and ensemble notation omitted for brevity):
\begin{equation}
  \begin{aligned}
    \vi_{t+\Delta t} &= \sigma( \W_i [\vxi_t, \vh_t] + \vb_i ), \\
    \vf_{t+\Delta t} &= \sigma( \W_f [\vxi_t, \vh_t] + \vb_f ), \\
    \vg_{t+\Delta t} &= \tanh( \W_g [\vxi_t, \vh_t] + \vb_g ), \\
    \vo_{t+\Delta t} &= \sigma( \W_o [\vxi_t, \vh_t] + \vb_o ), \\
    \vc_{t+\Delta t} &= \vf_{t+\Delta t} \odot \vc_t + \vi_{t+\Delta t} \odot \vg_{t+\Delta t}, \\
    \vh_{t+\Delta t} &= \vo_{t+\Delta t} \odot \tanh( \vc_{t+\Delta t} ),
  \end{aligned}
\end{equation}
where $\vi_t$, $\vf_t$, $\vg_t$, and $\vo_t$ are input, forget, cell, and output gates, respectively.
Matrices $\W_i$, $\W_f$, $\W_g$, $\W_o \in \RR^{d_h \times (d_h + d_z)}$, and vectors $\vb_i$, $\vb_f$, $\vb_g$, $\vb_o \in \RR^{d_h}$ are trainable parameters.
Square brackets $[\dots]$ denote concatenation, $\sigma$ the sigmoid function, and $\odot$ element-wise multiplication.

The input state $\vxi_t^{(1)}$ of the first layer is equal to $\vxi_t^{(1)} = [\vz_t, \vq_t, \vEF_t]$, where $\vz_t \in \RR^{d_z}$ is the latent state, $\vq_t \in \RR^{d_q}$ the quantities of interest, and $\vEF_t \in \RR^{d_F}$ the external forcing.
In the Van der Pol oscillator (VdP) study, $\vz_t = \vz(t) = [x(t), y(t)]$, and $\vEF_t = f(t) = \mu(t)$.
In the reaction-diffusion case study, $\vz_t$ is the output of the autoencoder, and $f_t = d(t)$.
In the CFD study, $\vz(t)$ is the output of the (multiresolution) autoencoder, $\vq(t) = \alpha_F \vFcyl(t)$ the scaled force on the cylinder, and $f(t)$ the normalized Reynolds number $f(t) = \widetilde{\Re}(t) = 2.4 \Re(t) / \Re_\text{max} - 1.2$.
In studies with $\Re \leq 1000$, $\Re_\text{max} = 1000$, and in studies with $\Re$ up to 1200, $\Re_\text{max} = 1200$.
For other layers $l \geq 2$, the input state $\vxi_t^{(l)}$ is equal to the previous layer's hidden state $\vh_{t+\Delta t}^{(l-1)}$.

The residuals $\Delta \vw(t) = \vw(t + \Delta t) - \vw(t)$, $\vw(t) = [\vz(t), \vq(t)]$ and uncertainties $\vsigma_{\Delta w}(t)$ are given by
\begin{equation}
  \label{eq:lstm-sigma-layer}
  \begin{aligned}
    \Delta \vw(t) &= \W_w \vh_{t+\Delta t}^{(L)} + \vb_w, \\
    \vsigma^2_{\Delta \vw}(t) &= \SoftPlus_\epsilon ( \W'_{\sigma} \CELU( \W_{\sigma} \vh_{t+\Delta t}^{(L)} + \vb_{\sigma} )  + \vb_{\sigma}'),
  \end{aligned}
\end{equation}
where $\vh^{(L)}$ is the hidden state of the final layer $L$.
Matrices
$\W_w \in \RR^{(d_z + d_q) \times d_h}$,
$\W_w \in \RR^{d_{\sigma} \times d_h}$ and
$\W'_{\sigma} \in \RR^{(d_z + d_q) \times d_{\sigma}}$, and
biases
$\vb_w, \vb'_{\sigma} \in \RR^{d_z + d_q}$ and
$\vb_{\sigma} \in \RR^{d_{\sigma}}$, with $d_{\sigma} = 100$,
are trainable parameters.
Functions
$\CELU(x) = \max(0, x) + \min(0, \exp(x) - 1)$
and
$\SoftPlus_\epsilon(x) = \log(1 + \exp(x)) + \epsilon$, with $\epsilon = 10^{-6}$, are nonlinearities.
The parameters for $\vsigma_{\Delta \vw}$ are trained separately from the rest, as explained in \cref{sec:uncertainty}.
Concretely, the parameters $\W_{\sigma}$, $\W'_{\sigma}$, $\vb_{\sigma}$ and $\vb'_{\sigma}$ constitute the trainable parameters $\vtheta_{\sigma}$, whereas other parameters constitute $\vtheta_{\mu}$.

In the ensemble, each LSTM network, augmented with the additional layers for computing $\Delta \vw(t)$ and $\vsigma^2_{\Delta w}(t)$, is trained separately with its own trainable parameters, starting from its own randomly initialized values.
The networks are trained using the Adam optimizer~\cite{kingma2014adam} with backpropagation through time~\cite{werbos1988generalization} to minimize the trajectory sum of per-state losses described in \cref{sec:uncertainty}.
Finally, in the VdP study, $L = 3$ layers with hidden state size of $d_h = 32$ (per layer) were used, amounting to $\num{8176}$ trainable parameters in total.
In the reaction-diffusion case study, $L = 2$ layers with $d_h = 64$ were used, amounting to \num{60308} parameters (for $d_z = 8$).
In the CFD case study, $L = 2$ layers with $d_h = 32$ were used, amounting to \num{20944} parameters (for $d_z = 16$).
}  %

\section{Details of the Van der Pol oscillator case study}
\label{sec:appendix-vdp}

\begin{figure}
  \centering
  \includegraphics[width=0.5\columnwidth]{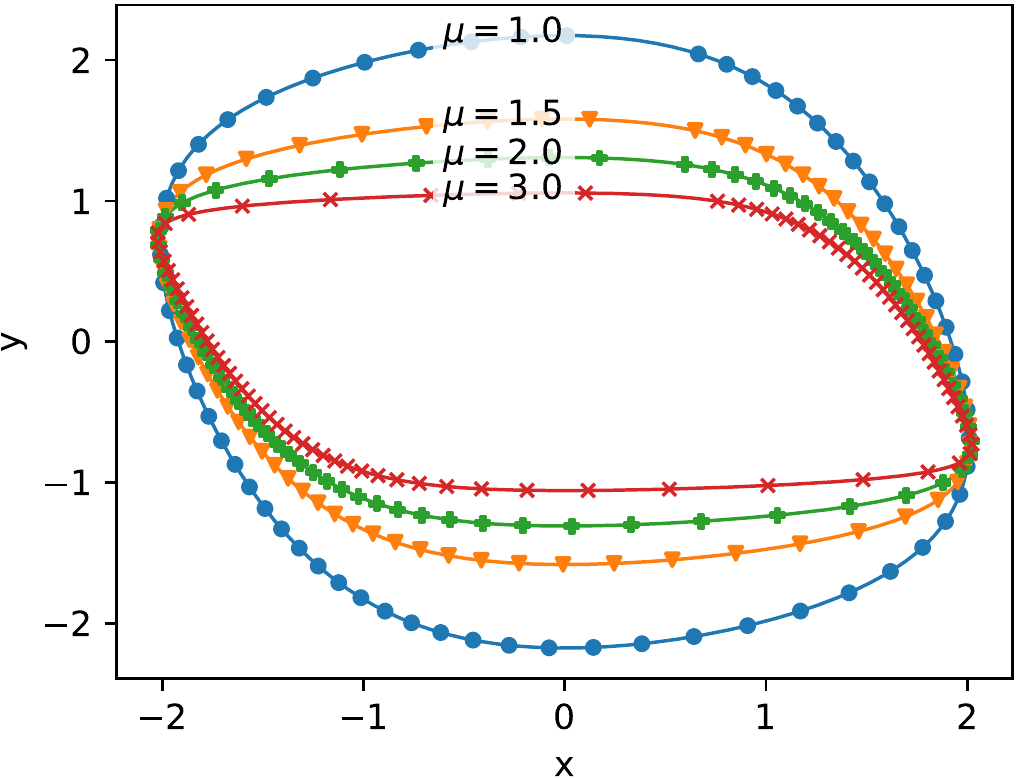}
  \caption{
    The limit cycles of the Van der Pol oscillator.
  }
  \label{fig:vdp-limit-cycles}
\end{figure}

The Van der Pol oscillator enters a limit cycle given a fixed $\mu$.
The limit cycles for different values of $\mu$ are shown in \cref{fig:vdp-limit-cycles}.

The details of the three cases of $\mu(t)$ are as follows.
In the $\muA(t)$ case, $\mu$ alternates between values $\mu = 1$ and $\mu = 3$ every \num{50000} time steps.
In the $\muB(t)$ case, $\mu$ changes between values 1.96, 1.23, 2.80, 2.34, 1.61, 2.57, 1.49, 3.00, 1.69, and 1.00, at the same rate as $\muA(t)$.
Finally, the $\muC(t)$ profile is computed by smoothing $\muB(t)$ and adding Brownian noise to it:
\begin{equation}
  \begin{aligned}
    \muC(t) &= \muB(t) + \alpha (\muC(t - \Delta t) - \muB(t)) + \beta \epsilon(t),
    \quad \epsilon(t) \sim \mathcal{U}([-1, 1]), \\
    \muC(0) &= \muB(0),
  \end{aligned}
\end{equation}
with $\alpha = \exp(-\Delta t / 200), \Delta t = 0.1$ and $\beta = 0.005$.

\subsection{Hyper-parameter study}
\label{sec:vdp-hyperparam-study}

The search space of hyper-parameters and their selected values are shown in \cref{tbl:vdp-hyperparam-study}.
The study was performed on 1536 samples of hyper-parameter values, randomly selected in the listed ranges, optimizing for the total macro utilization and average online validation error $E$.
Each simulation was run for \num{200000} time steps, with $\mu$ alternating between 1.5 and 2.5 every \num{25000} time steps.
Thresholds of $E_\text{max} = 0.14$ and $\sigma_\text{max} = 0.14$ were used.
The highest macro utilization achieved was $70\%$.
As the final hyper-parameter set (\cref{tbl:vdp-hyperparam-study}), we selected a Pareto-optimal sample that achieves $68\%$ macro utilization and average online validation error $E$ of $0.008$.
We further explored adversarial training~\cite{goodfellow2014explaining,lakshminarayanan2016simple}, but it did not affect the results noticeably.
Moreover, we tested how significantly better the network is with $\mu(t)$ as part of the input compared to not having access to $\mu(t)$.
The results show that the macro propagator the acceptance rate and the total utilization are still high (${\sim}54\%$) without providing $\mu(t)$ to the network.
However, naturally, in that case, the uncertainty of the macro propagator's prediction is insensitive to changes of $\mu(t)$ during the macro-only stage.

\begin{table}
  \caption{
    Van der Pol oscillator hyper-parameter study and final parameters in \textbf{bold}.
    Percentages on the right denote the highest achieved macro utilization.
  }
  \centering
  \begin{tabular}{c|cc}
    parameter & search space & comment \\
    \hline
    learning rate & $\operatorname{LogUniform}(0.0002, 0.02)$ & $\mathbf{0.002}$ \\
    batch size & $\{8, \mathbf{16}, 32, 64\}$ & \\
    LSTM hidden state size $d_h$ & $\{8, 16, \mathbf{32}\}$ & \\
    number of LSTM layers $L$ & $\{1, 2, \mathbf{3}, 4\}$ & \\
    adversarial training? & \{\textbf{no}, yes\} & no significant effect \\
    $\mu(t)$ as part of input? & \{no, \textbf{yes}\} &  ${\sim}54\%$ vs ${\sim}70\%$ \\
  \end{tabular}
  \label{tbl:vdp-hyperparam-study}
\end{table}

\subsection{Dependence of error on thresholds and ensemble size}
\label{sec:vdp-E-vs-sigma-max-K}

\begin{figure*}
  \begin{subfigure}{\textwidth}
    \centering
    \includegraphics[width=0.8\textwidth]{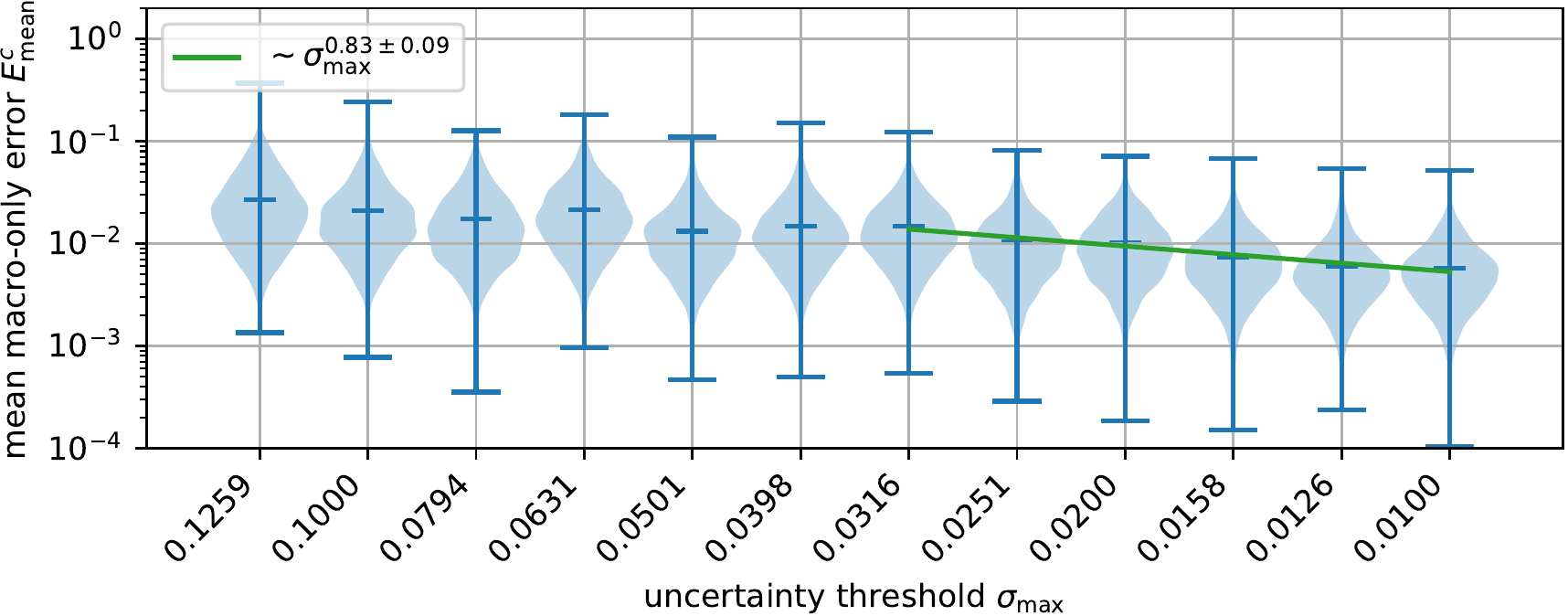}
  \end{subfigure}
  \medskip
  \begin{subfigure}{\textwidth}
    \centering
    \includegraphics[width=0.8\textwidth]{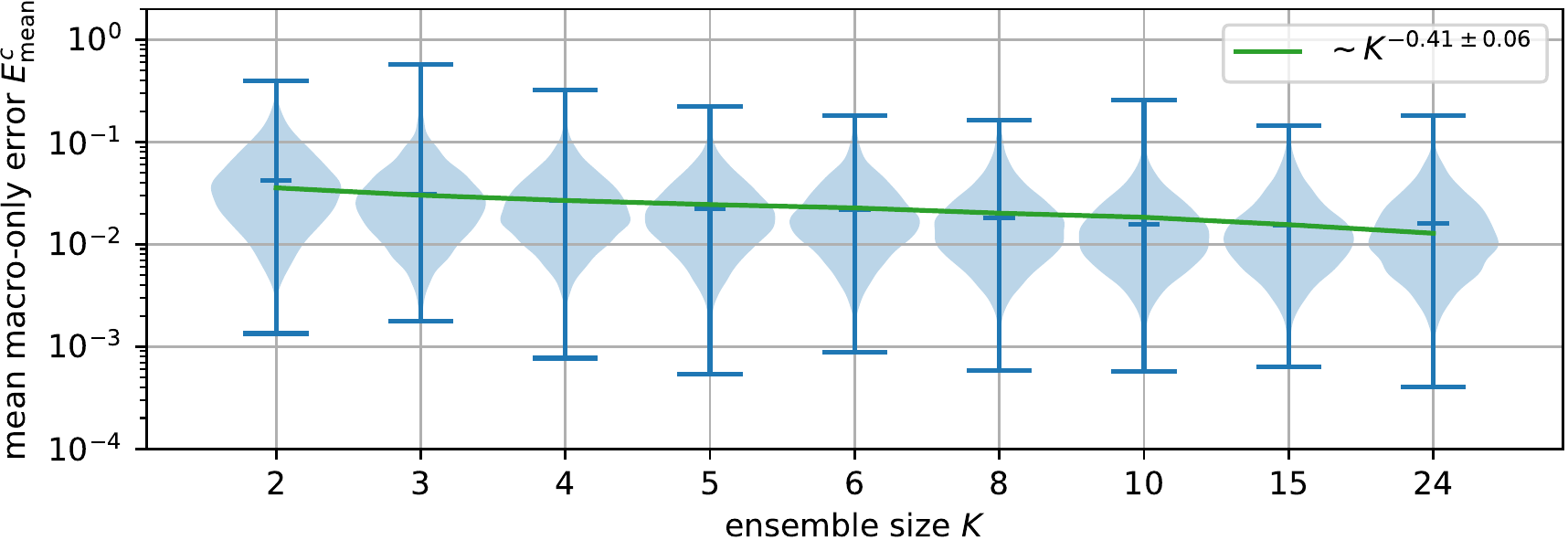}
  \end{subfigure}
  \caption{
    Dependence of the prediction error on the uncertainty threshold $\sigma_\text{max}$ (top) and the ensemble size $K$ (bottom), in the Van der Pol oscillator case study (\cref{sec:vdp-E-vs-sigma-max-K}).
    Each violin plot represents one run and shows the distribution of mean macro-only prediction errors along the AdaLED cycles.
  }
  \label{fig:vdp-Ecmean}
\end{figure*}

The testing error $E$ can be decreased with a stricter uncertainty threshold $\sigma_\text{max}$ or with a larger ensemble size $K$.
To analyze the extent of their effect on $E$, we run the $\muA(t)$ case for varying $\sigma_\text{max}$ and $K$.
The threshold $E_\text{max}$ is fixed to 0.1.
The top plot in \cref{fig:vdp-Ecmean} shows the distributions of mean cycle prediction errors $E^c_\text{max}$ (the average over macro-only steps of a cycle $c$) for varying $\sigma_\text{max}$ and $K$.
We notice that, for sufficiently small $\sigma_\text{max}$, the error drops approximately linearly with respect to $\sigma_\text{max}$.
This trend can be explained through dimensionality analysis.
Namely, $E$ and $\sigma_\text{max}$ are quantities of the same units.
The bottom plot of \cref{fig:vdp-Ecmean} shows that $E$ drops approximately as $1 / \sqrt{K}$, which is in accordance with the central limit theorem.
Concretely, if we assume that each individual LSTM produces the correct trajectory up to the noise of zero bias, then the noise cancels out at the rate of $1 / \sqrt{K}$.

It should be noted, however, that although stricter $\sigma_\text{max}$ improves error, it decreases macro utilization $\eta$.
For example, for $\sigma_\text{max} = 0.1$, $0.02$ and $0.01$, the macro utilization $\eta$ is $60\%$, $29\%$ and $18\%$, respectively.
Likewise, increasing the ensemble size from $K = 5$ to $K = 20$ reduces the error by ${\sim}$2x, but increases the total training time by 4x, assuming a fixed number of epochs.
Thus, depending on the situation and objectives, decreasing $\sigma_\text{max}$ and increasing $K$ may or may not be favorable.

\section{Details of the reaction-diffusion study}
\label{sec:appendix-rd}

The initial condition of the system is given by~\cite{champion2019data}
\begin{equation}
  \label{eq:rd-initial-conditions}
  \begin{aligned}
    u(x, y, 0) &= \tanh \left( \sqrt{x^2 + y^2} \cos \left( \operatorname{atan2}(y, x) - \sqrt{x^2 + y^2} \right) \right), \\
    v(x, y, 0) &= \tanh \left( \sqrt{x^2 + y^2} \sin \left( \operatorname{atan2}(y, x) - \sqrt{x^2 + y^2} \right) \right).
  \end{aligned}
\end{equation}

\Cref{eq:rd} is integrated in time using the fourth-order Runge-Kutta-Fehlberg integration scheme.
A second-order centered stencil with zero von Neumann boundary conditions is used for the diffusion term.

The capacity of the dataset is set to 1024 trajectories, each having 24 time steps.
The hyper-parameters, including the latent space dimension, RNN hidden state size, number of layers, batch size, learning rate, the amount of training per AdaLED cycle, and the thresholds $E_\text{max}$ and $\sigma_\text{max}$ are all hand-tuned.
The autoencoder is composed of an encoder and a decoder, each a 4-layer convolutional neural network with 16 channels per layer.
The autoencoder architecture is provided in \cref{tbl:rd-cnn-architecture}.
The values $(\vec{u}, \vec{v})$, which span the range $[-1, 1]$, are downscaled by a factor of $1.1$ before entering the encoder and upscaled back at the end of the decoder.
A learning rate of 0.001 and a batch size of 64 are used for both the autoencoder and RNNs.
The training is performed in partial epochs, with the autoencoder trained on $6.25\%$ of states in the dataset and the RNNs trained on $12.5\%$ of stored trajectories after each AdaLED cycle.

\begin{table*}
  \small
  \centering
  \caption{
    The architecture of the convolutional autoencoder for the reaction-diffusion case study.
    All convolutional layers use \texttt{\detokenize{padding_mode=replicate}}.
    Batch normalization layers use the default parameters from PyTorch.
  }
  \centering
  \begin{tabular}{r|c|c}
    ID & shape & layer \\
    \hline
      & $2 \times 96 \times 96$  & Input \\
    \hline
     1 & $16 \times 96 \times 96$ & \detokenize{Conv(2, 16, kernel_size=5, padding=2)} \\
     1 & $16 \times 96 \times 96$ & \detokenize{BatchNorm()} \\
     2 & $16 \times 48 \times 48$ & \detokenize{AvgPool(kernel_size=2, stride=2)} \\
     3 & $16 \times 48 \times 48$ & \detokenize{CELU()} \\
     4 & $16 \times 48 \times 48$ & \detokenize{Conv(16, 16, kernel_size=5, padding=2)} \\
     1 & $16 \times 48 \times 48$ & \detokenize{BatchNorm()} \\
     5 & $16 \times 24 \times 24$ & \detokenize{AvgPool(kernel_size=2, stride=2)} \\
     6 & $16 \times 24 \times 24$ & \detokenize{CELU()} \\
     7 & $16 \times 24 \times 24$ & \detokenize{Conv(16, 16, kernel_size=5, padding=2)} \\
     1 & $16 \times 24 \times 24$ & \detokenize{BatchNorm()} \\
     8 & $16 \times 12 \times 12$ & \detokenize{AvgPool(kernel_size=2, stride=2)} \\
     9 & $16 \times 12 \times 12$ & \detokenize{CELU()} \\
    13 & $16 \times 12 \times 12$ & \detokenize{Conv(16, 16, kernel_size=5, padding=2)} \\
    14 & $16 \times  6 \times  6$ & \detokenize{AvgPool(kernel_size=2, stride=2)} \\
    15 & $16 \times  6 \times  6$ & \detokenize{CELU()} \\
    16 & 576                      & \detokenize{Flatten()} \\
    17 & $d_z^{(1)} = 8$          & \detokenize{Linear()} \\
    18 & $d_z^{(1)} = 8$          & \detokenize{Tanh()} \\
    \hline
       & $d_z^{(1)} = 8$          & $\vz^{(i)}$ \\
    \hline
     1 & 576                      & \detokenize{Linear()} \\
     2 & $16 \times  6 \times  6$ & \detokenize{ViewLayer()} \\
     3 & $16 \times 12 \times 12$ & \detokenize{Upsample(scale_factor=2.0, mode=bilinear))} \\
     4 & $16 \times 12 \times 12$ & \detokenize{Conv(16, 16, kernel_size=3, padding=1)} \\
     1 & $16 \times 12 \times 12$ & \detokenize{BatchNorm()} \\
     8 & $16 \times 12 \times 12$ & \detokenize{CELU()} \\
     9 & $16 \times 24 \times 24$ & \detokenize{Upsample(scale_factor=2.0, mode=bilinear))} \\
    10 & $16 \times 24 \times 24$ & \detokenize{Conv(16, 16, kernel_size=5, padding=2)} \\
     1 & $16 \times 24 \times 24$ & \detokenize{BatchNorm()} \\
    11 & $16 \times 24 \times 24$ & \detokenize{CELU()} \\
    12 & $16 \times 48 \times 48$ & \detokenize{Upsample(scale_factor=2.0, mode=bilinear))} \\
    13 & $16 \times 48 \times 48$ & \detokenize{Conv(16, 16, kernel_size=5, padding=2)} \\
     1 & $16 \times 48 \times 48$ & \detokenize{BatchNorm()} \\
    14 & $16 \times 48 \times 48$ & \detokenize{CELU()} \\
    15 & $16 \times 96 \times 96$ & \detokenize{Upsample(scale_factor=2.0, mode=bilinear))} \\
    16 & $ 2 \times 96 \times 96$ & \detokenize{Conv(16, 2, kernel_size=5, padding=2)} \\
    17 & $ 2 \times 96 \times 96$ & \detokenize{Tanh()} \\
    \hline
    & 50K & total number of parameters \\
  \end{tabular}
  \label{tbl:rd-cnn-architecture}
\end{table*}

\section{Details of the flow behind the cylinder study}

To trigger vortex shedding, we add a short symmetry-breaking vertical movement at the start of the simulation:
\begin{equation}
  u^s_y(t) = e^{-\alpha t} \sin(\beta t) u_{y0}^s,
\end{equation}
where $\alpha = 100$, $\beta = 200$ and $u_{y0}^s = 0.05d$.

The simulations were performed using \cubismAMR{}~\cite{chatzimanolakis2022cubismamr}, an adaptive mesh refinement (AMR) CPU--GPU hybrid C++ code for solving the incompressible Naiver-Stokes equations.
To use it within AdaLED and from Python, we added Python bindings~\cite{jakob2017pybind11}, APIs for controlling the execution of the simulation, and APIs for exporting and importing the state.
The existing coarse-fine AMR interpolation schemes were reused for exporting and importing the state as a uniform grid.

The force $\vFcyl$ that the fluid exerts on the cylinder is given by the sum of the pressure and viscous forces and is provided by \cubismAMR{}:
\begin{equation}
  \label{eq:ns-force}
  \begin{aligned}
    \vFcyl &= \vF_\text{p} + \vF_\text{v}, \\
    \vF_\text{p} &= \oiint -p \vec{n} \dd{S}, \\
    \vF_\text{v} &= \nu \rho \oiint \left( \nabla \vu + \nabla \vu^\intercal \right) \cdot \vec{n} \dd{S},
  \end{aligned}
\end{equation}
where $S$ is the surface of the cylinder, and $\vec{n}$ the outward normal vector.

\subsection{Autoencoder for the CFD state}
\label{sec:cfd-autoencoder}

To achieve high speed-ups, the macro propagator is not operating on the high-dimensional micro state $\vu \in \RR^{d_v}$, $d_v = 512 \times 1024 \times 2 \approx 10^6$ directly, but on a smaller low-dimensional latent state $\vz \in \RR^{d_z}$, with $d_z \sim 10$.
The assumption is that this transition can indeed be performed: while we need high resolution to simulate the flow dynamics accurately and to acquire accurate forces $\vFcyl$, the actual dimensionality of the dynamics may be low.

To compress the velocity field $\vu$ to the latent state $\vz$, we use an autoencoder based on convolutional neural networks (CNNs).
Instead of training the autoencoder to naively reproduce $\vu$ by utilizing a simple (relative) MSE of $\vu$, we take into account the characteristics of the fluid dynamical system:
(i) physically essential quantities are also the spatial derivatives of the velocity (see~\cref{eq:ns-brinkman}) and the vorticity $\omega = (\nabla \times \vu)_z$,
(ii) the flow is incompressible, hence the divergence must be zero ($\nabla \cdot \vu = 0$),
(iii) we assume that the flow far from the cylinder requires smaller reconstruction accuracy compared to the flow around the cylinder.

If the autoencoder is trained only to minimize the MSE of $\vu$ while ignoring the value of derivatives, the reconstructed $\vu$ would have high spatial noise, resulting in inaccurate local derivatives and locally high vorticity $\omega = \omega_z = (\nabla \times \vu)_z$.
This noisy vorticity would cause unnecessary mesh refinement in the CFD solver used in this study~\cite{chatzimanolakis2022cubismamr}, which uses adaptive non-uniform mesh and magnitude of local vorticity as the mesh refinement and coarsening criterion.
We extend the loss function with a relative $L_1$ vorticity reconstruction error to ensure the reconstructed vorticity is low where it originally is low.
This helps reduce the mesh size by about 10-15\% compared to having no vorticity loss, and reduces the performance degradation that would partially cancel out the benefit of AdaLED.

Non-zero divergence $\nabla \cdot \vu$ can be prevented entirely as a hard constraint by predicting the stream function $\psi$ instead of the velocity field $\vu$~\cite{mohan2020embedding}.
The velocity field $\vu$ is then given as:
\begin{equation}
  \label{eq:stream-to-v}
  u_x = \pdv{\psi}{y}, \quad u_y = -\pdv{\psi}{x}.
\end{equation}

In its discretized form, the derivatives for computing $\vu$ from $\psi$ and $\omega$ from $\vu$ are computed using the 2\textsuperscript{nd} order accurate centered stencil.
For example, for a field $f$ and a grid spacing of $\Delta x$, the $x$-derivative is given as:
\begin{equation}
  \eval{\pdv{f}{x}}_{ij} = \frac{f_{i,j+1} - f_{i,j-1}}{2 \Delta x} + \mathcal{O}(\Delta x^2).
\end{equation}
Thus, taking derivatives removes one cell from each side of each dimension.
To account for that, $\psi$ is predicted with one cell of padding.

Finally, we want to prioritize reducing the reconstruction loss in the vicinity of the cylinder because this part affects the force $\vFcyl$ and because any error there will propagate to the rest of the flow.
Moreover, the vortex street behind the cylinder is relatively smooth and does not require high resolution.
We take advantage of these two observations and use a multiresolution autoencoder with two encoder--decoder pairs:
One for the whole domain downsampled to half the resolution ($\vu^{(1)}$), and one focusing on the subdomain around the cylinder at full resolution ($\vu^{(2)}$).
The two pairs operate independently and their compressed latent state $\vz^{(k)} \in \RR^{d_z^{(k)}}$ are concatenated into the final latent state $\vz \in \RR^{d_z}$, $d_z = d_z^{(1)} + d_z^{(2)}$.
The details are explained in the following section.

\subsection{Multiresolution autoencoders}
\label{sec:mr}

{  %

\newcommand{\vD}{\vec{D}}
\newcommand{\vU}{\vec{U}}
\newcommand{\vfoo}{\bm{\phi}}
\newcommand{\foo}{\bm{\phi}}
\newcommand{\Ny}{H}
\newcommand{\Nx}{W}
\newcommand{\loss}{\ell}
When building an autoencoder for 2D (or 3D) arrays, in cases where different parts of the flow exhibit different features and where not all parts require the same level of accuracy, we may benefit from combining multiple autoencoders operating at different spatial resolutions into a single one.
This enables us to reduce memory requirements and improve computational efficiency and training accuracy (accuracy is positively affected by improved processing speed and potentially from the benefits of a specialized architecture).
This section describes how such \emph{multiresolution autoencoders} can be constructed.
For simplicity, we focus on autoencoders reconstructing a 2D scalar array $\vfoo \in \RR^{\Ny \times \Nx}$ using two encoder--decoder pairs.
The method can be easily generalized to vector arrays, to more than two encoder--decoder pairs, and higher-dimensional arrays.

For each encoder--decoder pair $\operatorname{AE}_k$, $k \in \{1, 2\}$, we define a downsampling operation $\vD^{(k)} : \RR^{\Ny \times \Nx} \to \RR^{\Ny^{(k)} \times \Nx^{(k)}}$ that converts the full-resolution array $\vfoo$ into a smaller array $\vfoo^{(k)} = \vD^{(k)}(\vfoo)$ that $\operatorname{AE}_k$ will operate on.
In this case study, $\operatorname{AE}_1$ is used to reconstruct the whole domain at half the resolution ($\Ny^{(1)} = \Ny/2, \Nx^{(1)} = \Nx/2$), whereas $\operatorname{AE}_2$ is used for the detailed part of some size $\Ny^{(2)} \times\Nx^{(2)}$ around the cylinder.
Functions $\vD^{(1)}$ and $\vD^{(2)}$ are thus given as:
\begin{equation}
  \begin{aligned}
    D^{(1)}_{ij}(\vfoo) &= \frac{1}{4}\left( \foo_{2i,2j} + \foo_{2i,2j+1} + \foo_{2i+1,2j} + \foo_{2i+1,2j+1} \right),
    \quad & 0 \leq i < H^{(1)}, 0 \leq j < W^{(1)} \\
    D^{(2)}_{ij}(\vfoo) &= \foo_{i_0 + i, j_0 + j},
    \quad & 0 \leq i < H^{(2)}, 0 \leq j < W^{(2)} \\
  \end{aligned}
\end{equation}
where $i_0$ and $j_0$ are offsets of $\vfoo^{(2)}$ with respect to $\vfoo$ (indexing is 0-based).

The total reconstruction loss is defined as a weighted sum of the reconstruction losses of each individual $\operatorname{AE}_k$:
\begin{equation}
  \label{eq:mr-total-loss}
  \loss(\tilde{\vfoo}, \vfoo) = w^{(1)} \loss^{(1)}(\tilde{\vfoo}^{(1)}, \vfoo^{(1)}) + w^{(2)} \loss^{(2)}(\tilde{\vfoo}^{(2)}, \vfoo^{(2)}),
\end{equation}
where $w^{(k)}$ are weight factors.
Since the $\operatorname{AE}_k$s are independent, the weights $w^{(k)}$ effectively determine the relative learning rate between them.
For simplicity, we take $w^{(1)} = w^{(2)} = 1$.

The individual losses $l^{(k)}$ take into consideration that we do not want to waste the limited expressiveness of $\operatorname{AE}_1$ on reconstructing the part that $\operatorname{AE}_2$ is already focusing on.
Furthermore, to reduce the boundary effects (at which the reconstruction might be poor), we also want to exclude the edges from the reconstruction loss of $\operatorname{AE}_2$.
We, thus, use weighted (relative) reconstruction losses:
\begin{equation*}
  \loss^{(k)} \left( \tilde{\vfoo}^{(k)}, \vfoo^{(k)} \right)
    = \frac{ \sum_{ij} \alpha^{(k)}_{ij} \left( \tilde{\vfoo}^{(k)}_{ij} - \vfoo^{(k)}_{ij} \right)^2 }
           { \sum_{ij} \alpha^{(k)}_{ij} \left( \vfoo^{(k)}_{ij} \right)^2 + W^{(k)} H^{(k)} \epsilon_\phi }
\end{equation*}
for the relative MSE loss, or
\begin{equation*}
  \loss^{(k)} \left( \tilde{\vfoo}^{(k)}, \vfoo^{(k)} \right)
    = \frac{ \sum_{ij} \alpha^{(k)}_{ij} \abs{ \tilde{\vfoo}^{(k)}_{ij} - \vfoo^{(k)}_{ij} } }
           { \sum_{ij} \alpha^{(k)}_{ij} \abs{ \vfoo^{(k)}_{ij} } + W^{(k)} H^{(k)} \epsilon_\phi }
\end{equation*}
for the relative $L_1$ loss.
Here, field $\tilde{\vfoo}^{(k)}$ denotes the autoencoder reconstruction, $\vfoo^{(k)}$ the input and the target, and $\epsilon_\phi > 0$ a normalization offset for preventing diverging gradients.
The weight factors $\valpha^{(k)}$ are selected such that the center of the cylinder does not affect the loss of $\operatorname{AE}_1$ and that the edge of the second subdomain does not affect the loss of $\operatorname{AE}_2$:
\begin{equation}
  \begin{aligned}
    \valpha^{(1)} &= 1 - \vD^{(1)}(\vec{S}(d^{(1)})), \\
    \valpha^{(2)} &= \vD^{(2)}(\vec{S}(d^{(2)})), \\
  \end{aligned}
\end{equation}
where $\vec{S} : \RR \to \RR^{\Ny \times \Nx}$ is a 2D smoothed rectangular function:
\begin{equation}
  \begin{aligned}
    S_{ij}(d) &= S \left( \frac{\min\{ i' - i_0, i_1 - i' \} - d}{s} \right) \\
        & \times S \left( \frac{\min\{ j' - j_0, j_1 - j' \} - d}{s} \right), \\
    i' &= i + \frac{1}{2}, \quad \quad \text{(for cell-centered values)} \\
    j' &= j + \frac{1}{2}, & \\
    S(x) &= \frac{1}{1 + e^{-x}},  \quad \text{(sigmoid)}
  \end{aligned}
\end{equation}
where $(i_0, j_0)$ and $(i_1, j_1) = (i_0 + \Ny^{(2)}, j_0 + \Nx^{(2)})$ are the start and the end bounds of $\operatorname{AE}_2$.
Parameter $s > 0$ is a smoothing factor, and $d^{(1)}, d^{(2)} > 0$ the spatial margins.

The final reconstruction of $\vfoo \in \RR^{\Ny \times \Nx}$ from $\vfoo^{(1)}$ and $\vfoo^{(2)}$ is as well done in a weighted manner:
\begin{equation}
  \label{eq:mr-reconstruction}
  \begin{aligned}
    \vfoo &= \vbeta^{(1)} \odot \vU( \vfoo^{(1)} ) + \vbeta^{(2)} \odot \vU( \vfoo^{(2)} ), \\
    \vbeta^{(1)} &= 1 - \vec{S}(d^r) \\
    \vbeta^{(2)} &= \vec{S}(d^r) \\
  \end{aligned}
\end{equation}
where $\vU^{(k)} : \RR^{\Ny^{(k)} \times \Nx^{(k)}} \to \RR^{\Ny \times \Nx}$ are upsampling operations, $d^r$ the reconstruction margin, and operator $\odot$ the element-wise multiplication.
In this case study, $\vU^{(1)}$ is the upsampling operation with bilinear interpolation and $\vU^{(2)}$ is a zero-padding operation.

We tested two variants of $\operatorname{AE}_2$ that operate on different resolutions: $256 \times 256$ and $224 \times 224$.
As shown in \cref{sec:mr-study}, the latter variant exhibited slightly better performance for large macro utilizations and was selected as part of the reference parameter set.
The parameters are shown in \cref{tbl:mr-params}, and the geometry is visualized in \cref{fig:mr-geometry}.
By using margins $d^{(1)} > d^r > d^{(2)}$, we ensure that both $\operatorname{AE}_1$ and $\operatorname{AE}_2$ accurately reconstruct the part where the smoothed blending occurs ($0 < \beta_{ij}^{(k)} < 1$).

The multiresolution approach decreases the memory and storage requirements by 2.9x (for the $224 \times 224$ variant) and accelerates the training by approximately the same factor at a small cost of accuracy degradation.
Concretely, the relative mean square error between the original velocity field $\vu$ and the downsampled--upsampled $\vu'$ is $\norm{\vu' - \vu}_2^2 / \norm{\vu}_2^2 \approx 10^{-6}$.

\begin{figure}
  \centering
  \includegraphics[width=0.6\columnwidth]{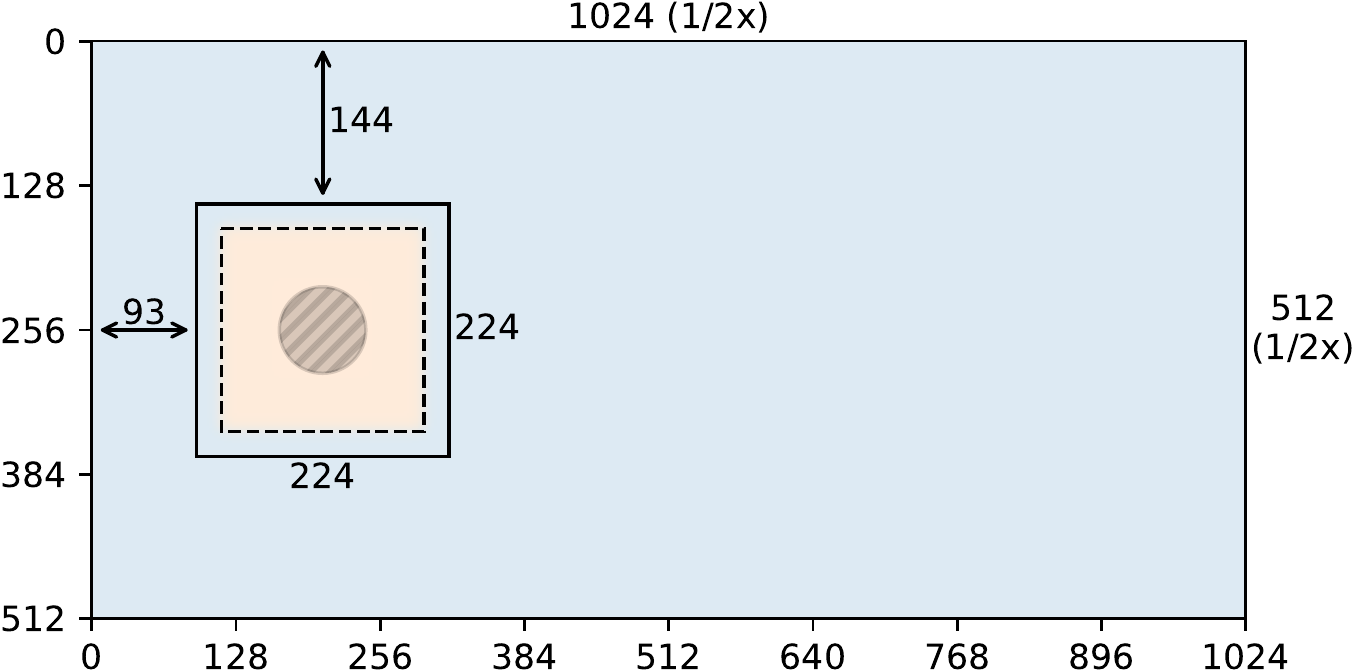}
  \caption{
    The geometry of the multiresolution autoencoder with resolution of $\operatorname{AE}_2$  equal to $224 \times 224$.
    The solid inner box represents the subdomain that $\operatorname{AE}_2$ operates on,
    the dashed line the distance $d^r = 22$ from the inner subdomain boundary (where $\vbeta^{(1)} = \vbeta^{(2)} = 0.5$).
    Colors represent the blending contributions ($\vbeta^{(1)} > 0.5$ marked as blue, $\vbeta^{(2)} > 0.5$ with orange),
    and circle the cylinder.
  }
  \label{fig:mr-geometry}
\end{figure}

\begin{table}
  \caption{Geometry and loss function parameters for multiresolution autoencoders, and memory usage per single $\vu$ field in single precision.}
  \centering
  \begin{tabular}{c|cc}
    parameter & \textbf{variant 1} & variant 2 \\
    \hline
    original resolution & \multicolumn{2}{c}{$1024 \times 512$} \\
    $\operatorname{AE}_1$ resolution & \multicolumn{2}{c}{$512 \times 256$} \\
    $\operatorname{AE}_2$ resolution & $256 \times 256$ & $\mathbf{224 \times 224}$ \\
    $\operatorname{AE}_2$ begin & (77, 128) & (93, 144) \\
    $\operatorname{AE}_2$ end   & (333, 384) & (317, 368) \\
    weight $w^{(1)}$ & \multicolumn{2}{c}{1.0} \\
    weight $w^{(2)}$ & \multicolumn{2}{c}{1.0} \\
    margin $d^{(1)}$ & \multicolumn{2}{c}{30.0} \\
    margin $d^r$     & \multicolumn{2}{c}{22.0} \\
    margin $d^{(2)}$ & \multicolumn{2}{c}{14.0} \\
    smoothing $s$    & \multicolumn{2}{c}{3.0} \\
    \hline
    original size of $\vu$ & \multicolumn{2}{c}{\SI{4.00}{MB}} \\
    reduced size of $\vu$ & \SI{1.50}{MB} & \SI{1.38}{MB} \\
  \end{tabular}
  \label{tbl:mr-params}
\end{table}

}  %

\subsection{Autoencoder summary and loss function}

\begin{figure}
  \centering
  \includegraphics[width=\columnwidth]{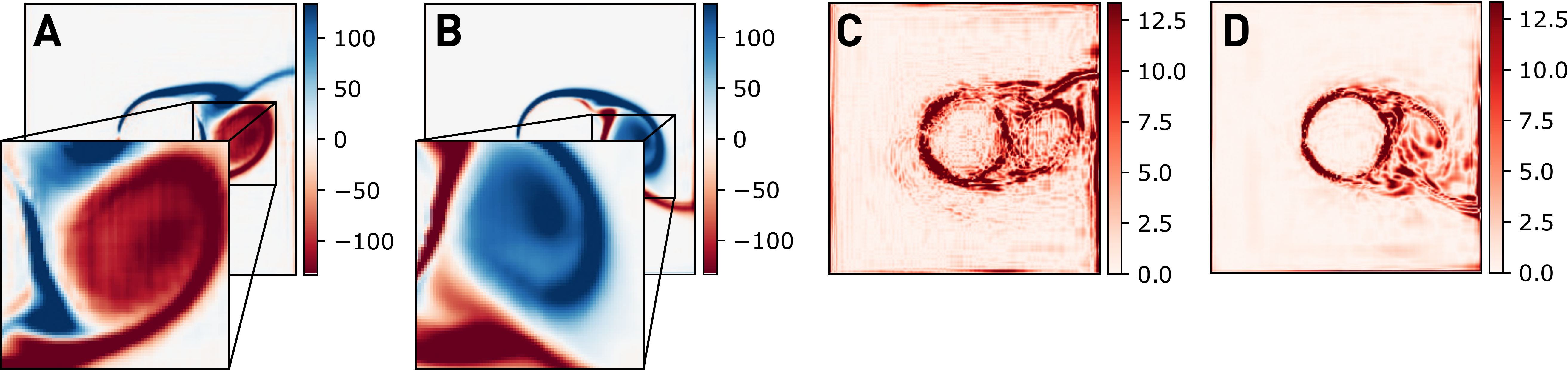}
  \caption{
    Reconstructed vorticity field (A and B) and its absolute error (C and D) for the $\operatorname{AE}_2$, for $\lambda_\omega = 0$ (without vorticity loss, A and C) and $\lambda_\omega = 0.03$ (with vorticity loss, B and D).
  }
  \label{fig:vorticity-loss}
\end{figure}

The discussion above is summarized in the following.
The state $\vu$ is stored in the dataset in two downsampled versions $\vu^{(1)}$ and $\vu^{(2)}$, each handled by its own encoder--decoder pair $\operatorname{AE}_k$.
For each $\operatorname{AE}_k$, $k \in \{1, 2\}$, the encoder $k$ takes the downsampled velocity field $\vu^{(k)} \in \RR^{H^{(k)} \times W^{(k)} \times 2}$ as input and compresses it into a latent state $\vz^{(k)} \in \RR^{d_z^{(k)}}$.
The decoder $k$ takes the latent state $\vz^{(k)}$ and decompresses it into the scalar field $\tilde{\psi}^{(k)}$ (with one cell of padding on each side of each dimension).
Then, the reconstructed velocity $\tilde{\vu}^{(k)}$ is computed from $\tilde{\psi}^{(k)}$ using \cref{eq:stream-to-v}.
The reconstruction loss of $\operatorname{AE}_k$ is defined as a weighted sum of the relative MSE loss of $\vu^{(k)}$ and the relative $L_1$ loss of vorticity $\vomega^{(k)}$ (notation $(k)$ omitted in the following for brevity):
\begin{align}
  \label{eq:cfd-reconstruction-pair-loss}
  \nonumber
  \ell^{(k)}(\tilde{\vu}, \vu)
      &= \lambda_u \frac{\sum_{ij} \alpha_{ij} (\tilde{\vu}_{ij} - \vu_{ij})^2}{\sum_{ij} \alpha_{ij} \vu_{ij}^2 + W^{(k)} H^{(k)} \epsilon_u} \\
      &+ \lambda_\omega \frac{\sum_{ij} \alpha'_{ij} \norm{\tilde{\vomega}_{ij} - \vomega_{ij}}_1}{\sum_{ij}{\alpha'_{ij} \norm{\vomega_{ij}}_1} + (W^{(k)} - 2)(H^{(k)} - 2) \epsilon_\omega}, \\
  \nonumber
  \tilde{\vomega} &= (\nabla \times \tilde{\vu})_z,
  \quad \vomega = (\nabla \times \vu)_z, \\
  \valpha' &= \valpha_{1..H-2;1..W-2} \in \RR^{(H - 2) \times (W - 2)},
\end{align}
where $\lambda_u = 1$ and $\lambda_\omega = 0.03$ are the weight factors,
and $\epsilon_u = 0.01$ and $\epsilon_\omega = 0.7$ the normalization offsets used to avoid exploding gradients when training on initial states where $\vu \approx \vec{0}$.
The effect of $\lambda_\omega$ is visualized in \cref{fig:vorticity-loss}.
Numbers $\epsilon_u$ and $\epsilon_\omega$ were selected to match ${\approx}25\%$ of the mean $(\vu_{ij}^{(1)})^2$ and the mean $\abs{\omega_{ij}^{(1)}}$, respectively, for the developed flow at $\Re = 500$.
The total loss $\ell(\dots)$ is defined as the weighted sum of the losses of $\operatorname{AE}_k$s:
\begin{equation}
  \label{eq:cfd-reconstruction-loss}
  \ell(\tilde{\vu}^{(1)}, \tilde{\vu}^{(2)}, \vu^{(1)}, \vu^{(2)}) = w^{(1)} \ell^{(1)}(\tilde{\vu}^{(1)}, \vu^{(1)}) + w^{(2)} \ell^{(2)}(\tilde{\vu}^{(2)}, \vu^{(2)}),
\end{equation}
where $w^{(1)} = w^{(2)} = 1$ are relative weights between $\operatorname{AE}_k$s.
Either when computing the online validation error $E$ in \cref{eq:cup-E} or when performing macro-to-micro transition, the full resolution velocity $\vu$ is reconstructed by merging $\vu^{(1)}$ and $\vu^{(2)}$ as described in \cref{eq:mr-reconstruction}.
The merging must be performed on velocities $\vu$ and not on the stream function $\psi$.
This is because the stream functions are defined up to an unspecified additive constant, making their merging impossible.
Furthermore, by smoothly blending between two upscaled velocity fields (\cref{eq:mr-reconstruction}), we ensure the spatial derivatives of $\vu$ are smooth.

\subsection{Hyper-parameters, the CNN architecture and training}
\label{sec:cup-hyperparam-and-cnn}

Apart from the hyper-parameters listed in the autoencoder study in \cref{sec:cup-training-study}, other parameters, such as the batch size, were hand-tuned and are listed in \cref{tbl:cup-hyper-other}.

The basis of the $\vu$ autoencoder are two convolutional autoencoders, each operating on one downsampled array $\vu^{(k)}$.
The two autoencoders share the same architecture but are trained separately.
Their final CNN architecture after the hyper-parameter study is shown in \cref{tbl:cnn-architecture}.

\begin{figure}
  \centering
  \includegraphics[width=0.3\columnwidth]{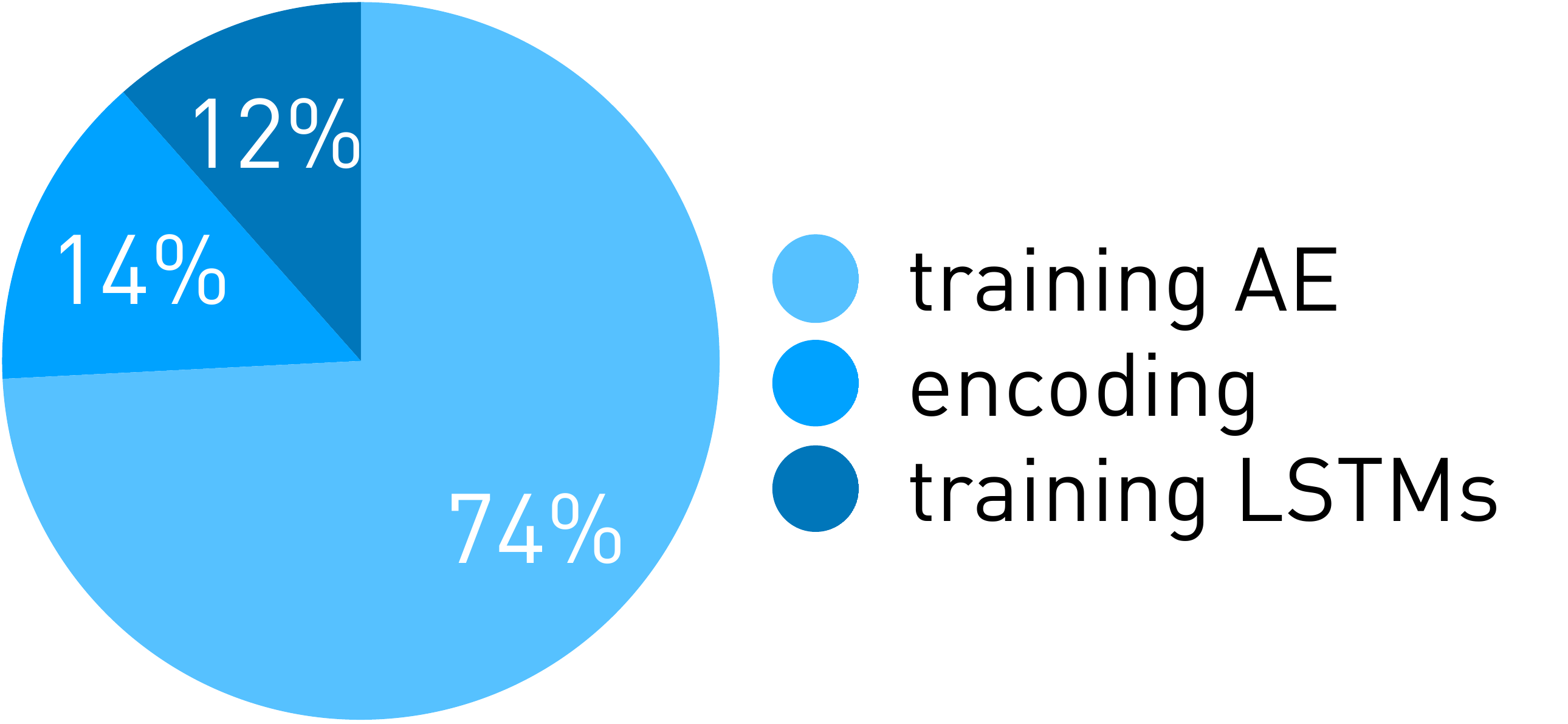}
  \caption{
    Fraction of the execution time of stages of the training in the flow behind the cylinder case.
  }
  \label{fig:training-time}
\end{figure}

The training is performed continuously in parallel with the simulation and AdaLED inference.
In each epoch, the networks are trained on $12.5\%$ of the dataset.
An epoch consists of training the autoencoder, encoding the states to build a temporary dataset for LSTMs, and finally, training the LSTMs.
The relative execution time of the three training stages is shown in \cref{fig:training-time}.

\begin{table}
  \caption{Hand-tuned network hyper-parameters for the flow behind the cylinder study. Other parameters are listed in \cref{sec:mr-study}.}
  \centering
  \begin{tabular}{c|c}
    hyper-parameter & value \\
    \hline
    autoencoder batch size & 8 \\
    LSTM batch size & 8 \\
    LSTM hidden state size & 32 \\
    number of LSTM layers & 2 \\
    single vs double precision & single \\
  \end{tabular}
  \label{tbl:cup-hyper-other}
\end{table}

\begin{table*}
  \small
  \centering
  \caption{
    The architecture of the convolutional autoencoders for the flow behind the cylinder case study.
    All convolutional layers use \texttt{\detokenize{padding_mode=replicate}}.
  }
  \centering
  \begin{tabular}{r|c|c|c}
    ID & AE \#1 shape & AE \#2 shape & layer \\
    \hline
      & $2 \times 256 \times 512$ & $2 \times 224 \times 224$ & Input \\
    \hline
     1 & $16 \times 256 \times 512$ & $16 \times 224 \times 224$ & \detokenize{Conv(2, 16, kernel_size=5, padding=2)} \\
     2 & $16 \times 128 \times 256$ & $16 \times 112 \times 112$ & \detokenize{AvgPool(kernel_size=2, stride=2)} \\
     3 & $16 \times 128 \times 256$ & $16 \times 112 \times 112$ & \detokenize{CELU()} \\
     4 & $16 \times 128 \times 256$ & $16 \times 112 \times 112$ & \detokenize{Conv(16, 16, kernel_size=5, padding=2)} \\
     5 & $16 \times  64 \times 128$ & $16 \times  56 \times  56$ & \detokenize{AvgPool(kernel_size=2, stride=2)} \\
     6 & $16 \times  64 \times 128$ & $16 \times  56 \times  56$ & \detokenize{CELU()} \\
     7 & $16 \times  64 \times 128$ & $16 \times  56 \times  56$ & \detokenize{Conv(16, 16, kernel_size=5, padding=2)} \\
     8 & $16 \times  32 \times  64$ & $16 \times  28 \times  28$ & \detokenize{AvgPool(kernel_size=2, stride=2)} \\
     9 & $16 \times  32 \times  64$ & $16 \times  28 \times  28$ & \detokenize{CELU()} \\
    13 & $16 \times  32 \times  64$ & $16 \times  28 \times  28$ & \detokenize{Conv(16, 16, kernel_size=3, padding=1)} \\
    14 & $16 \times  16 \times  32$ & $16 \times  14 \times  14$ & \detokenize{AvgPool(kernel_size=2, stride=2)} \\
    15 & $16 \times  16 \times  32$ & $16 \times  14 \times  14$ & \detokenize{CELU()} \\
    16 & 8192                       & 3136                       & \detokenize{Flatten()} \\
    17 & $d_z^{(1)} = 8$            & $d_z^{(2)} = 8$            & \detokenize{Linear()} \\
    18 & $d_z^{(1)} = 8$            & $d_z^{(2)} = 8$            & \detokenize{Tanh()} \\
    \hline
       & $d_z^{(1)} = 8$            & $d_z^{(2)} = 8$            & $\vz^{(i)}$ \\
    \hline
     1 & 8192                       & 3136                       & \detokenize{Linear()} \\
     2 & $16 \times  16 \times  32$ & $16 \times  14 \times  14$ & \detokenize{ViewLayer()} \\
     3 & $16 \times  32 \times  64$ & $16 \times  28 \times  28$ & \detokenize{Upsample(scale_factor=2.0, mode=bilinear))} \\
     4 & $16 \times  32 \times  64$ & $16 \times  28 \times  28$ & \detokenize{Conv(16, 16, kernel_size=3, padding=1)} \\
     8 & $16 \times  32 \times  64$ & $16 \times  28 \times  28$ & \detokenize{CELU()} \\
     9 & $16 \times  64 \times 128$ & $16 \times  56 \times  56$ & \detokenize{Upsample(scale_factor=2.0, mode=bilinear))} \\
    10 & $16 \times  64 \times 128$ & $16 \times  56 \times  56$ & \detokenize{Conv(16, 16, kernel_size=5, padding=2)} \\
    11 & $16 \times  64 \times 128$ & $16 \times  56 \times  56$ & \detokenize{CELU()} \\
    12 & $16 \times 128 \times 256$ & $16 \times 112 \times 112$ & \detokenize{Upsample(scale_factor=2.0, mode=bilinear))} \\
    13 & $16 \times 128 \times 256$ & $16 \times 112 \times 112$ & \detokenize{Conv(16, 16, kernel_size=5, padding=2)} \\
    14 & $16 \times 128 \times 256$ & $16 \times 112 \times 112$ & \detokenize{CELU()} \\
    15 & $16 \times 256 \times 512$ & $16 \times 224 \times 224$ & \detokenize{Upsample(scale_factor=2.0, mode=bilinear))} \\
    16 & $ 1 \times 258 \times 514$ & $ 1 \times 224 \times 224$ & \detokenize{Conv(16, 1, kernel_size=5, padding=3)} \\
    17 & $ 2 \times 256 \times 512$ & $ 2 \times 224 \times 224$ & \detokenize{StreamFnToVelocity()} (\cref{eq:stream-to-v}) \\
    \hline
    & 171K & 85K & total number of parameters \\
  \end{tabular}
  \label{tbl:cnn-architecture}
\end{table*}

\subsection{Latent trajectory}
\label{app:cyl-latent}

\begin{figure}
  \centering
  \includegraphics[width=\columnwidth]{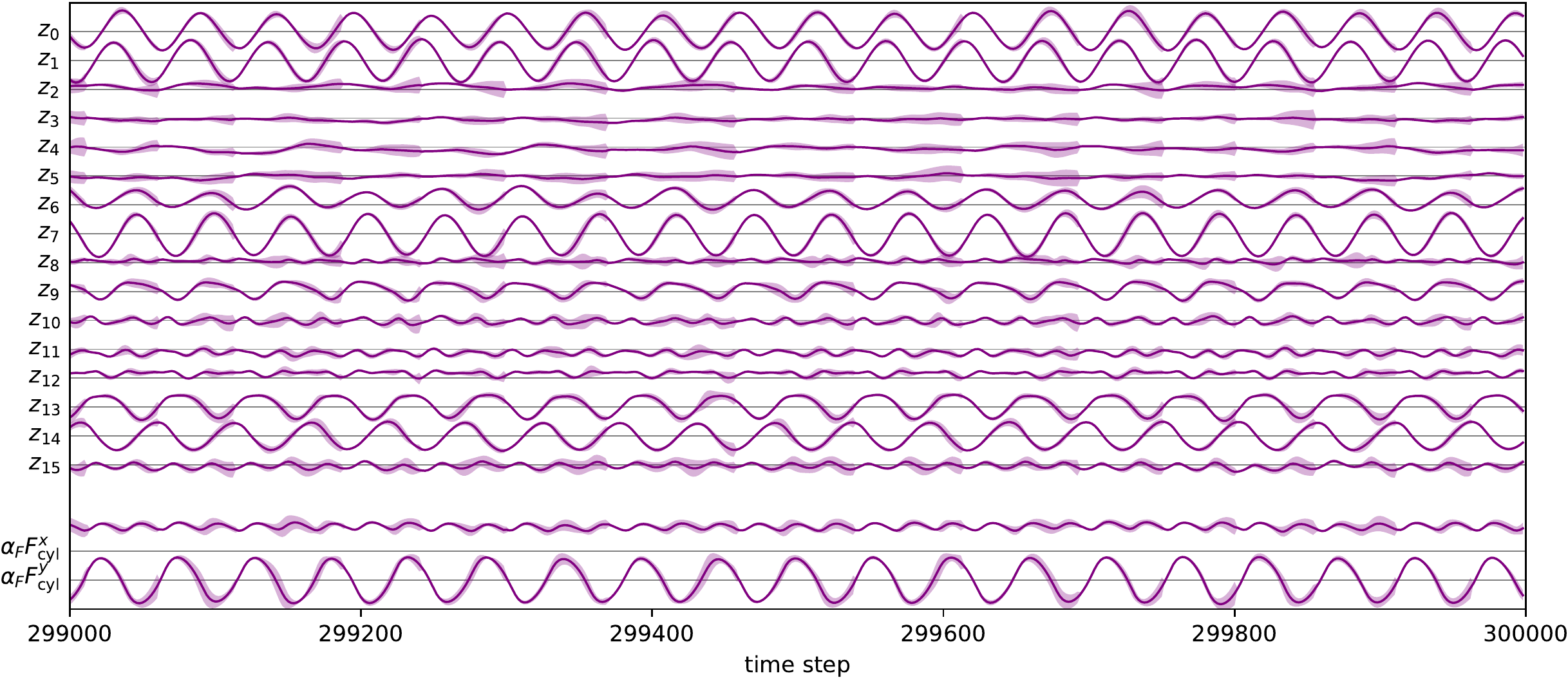}
  \caption{
    The trajectory $\vz(t)$ and $\alpha_F \vFcyl(t)$ from the simulation from \cref{sec:cup-selected}, as predicted by the ensemble.
    The solid line represents the ensemble mean prediction, and the faded region is the prediction uncertainty (the ensemble's standard deviation).
    For clarity, the uncertainties are enhanced by 8x.
    The numbers range between approx.\@ ${-}0.5$ and $0.5$.
    See \cref{app:cyl-latent}.
  }
  \label{fig:selected-latent-F}
\end{figure}

A section of the macro trajectory from the simulation from \cref{sec:cup-selected} is shown in \cref{fig:selected-latent-F}.
The first 16 lines correspond to the latent states $\vz(t)$ and the last two to the force $\vFcyl(t)$ (scaled with a factor of $\alpha_F = 7.2$).
The total uncertainty $\sigma$ is defined as the root square mean of all 18 uncertainties.
It can be seen that the majority of the uncertainty comes from low-amplitude latent variables.
The possibility of using weighted uncertainties, depending on the importance of each variable, is a topic of future research.

\subsection{Generalization to other Reynolds number profiles}
\label{app:cyl-Re-generalization}

\begin{figure}
  \centering
  \includegraphics[width=\columnwidth]{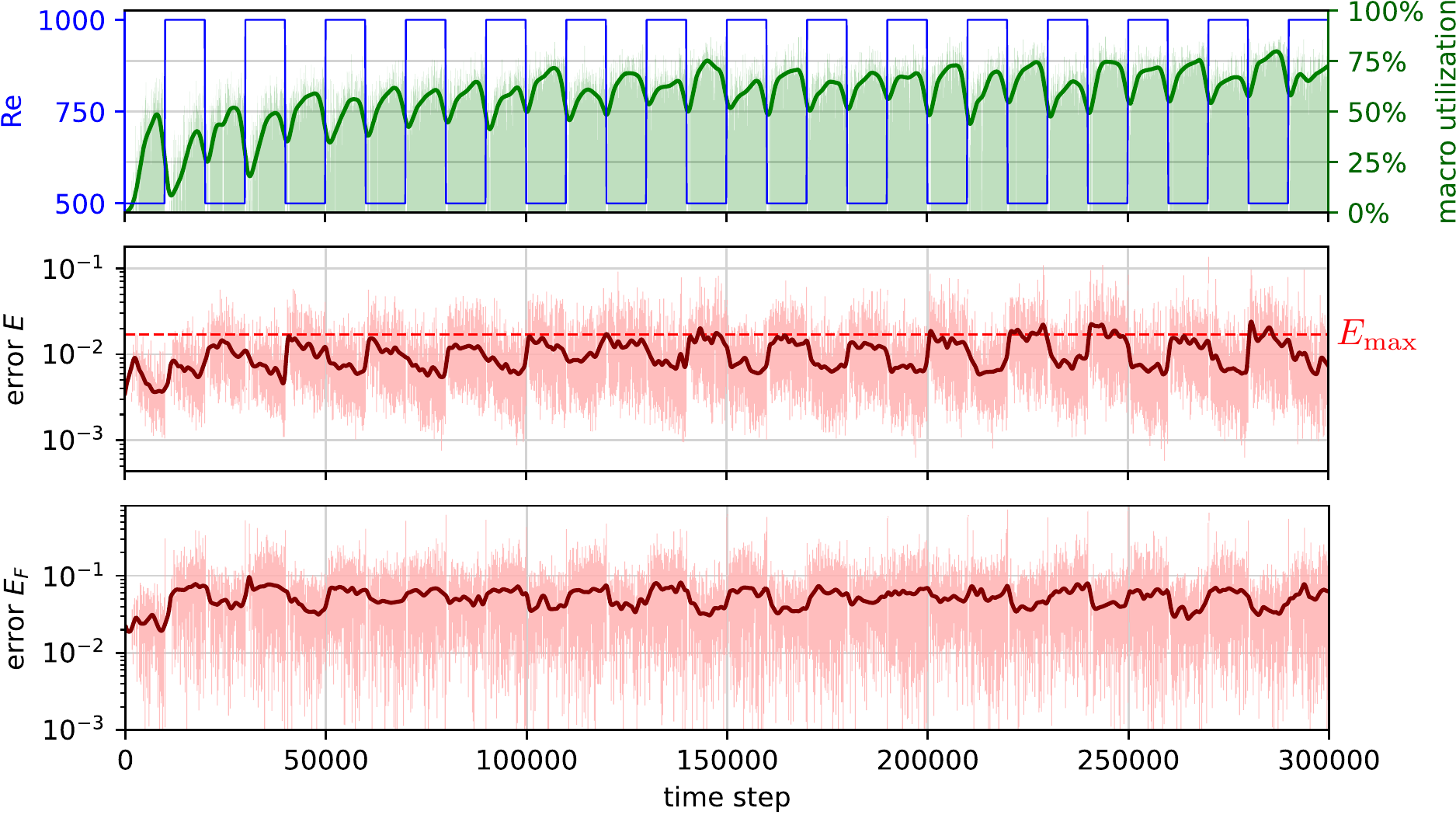}
  \caption{
    AdaLED performance on a flow behind cylinder simulation for $\Re(t) \in \{500, 1000\}$, analogous to \cref{fig:selected-utilization}.
    Top: Reynolds number $\Re(t)$ profile and the macro utilization $\eta$.
    Middle and bottom: validation errors of the velocity ($E$, \cref{eq:cup-E}) and force on the cylinder ($E_F$, \cref{eq:cup-EF}).
    The per-step errors (faded red) alternate between low values at the beginning of the macro-only stage and higher errors at the end of the macro-only stage.
  }
  \label{fig:generalization-performance}
\end{figure}

The hyper-parameters and thresholds used in simulations reported in \cref{sec:cup-selected} were fine-tuned for that specific Reynolds number profile of cycling between $\Re=600$, 750 and 900, updated every \num{5000} time steps, as described in \cref{sec:mr-study} (the hyper-parameter study used shorter simulations than the production runs).
To test the generalization of hyper-parameters and thresholds to another Reynolds number profile, we simulate with Reynolds number alternating between 500 and 1000 every \num{10000} time steps.
The macro utilization $\eta$, velocity field error $E$, and the cylinder force error $E_F$ are shown in \cref{fig:generalization-performance}.
Compared to the macro utilization of $\eta = 69\%$ (speed-up of 2.9x) in \cref{sec:cup-selected}, here, the achieved utilization is $58\%$ (speed-up of 2.1x).
As before, the average velocity and cylinder force errors $E$ and $E_F$ are $1\%$ and $5\%$.
In this case, changing the setup resulted in smaller speed-ups.
Thus, to achieve optimal performance, the hyper-parameters (particularly learning rates and thresholds) may have to be additionally fine-tuned if the simulation setup is updated.

\end{document}